\documentclass{article}

\begin{document}
\newcommand{\bq}{\begin{equation}}
\newcommand{\eq}{\end{equation}}
\newcommand{\bqa}{\begin{eqnarray}}
\newcommand{\eqa}{\end{eqnarray}}
\newcommand{\nl}{\nonumber \\}
\newcommand{\eqn}[1]{Eq.(\ref{#1})}
\newcommand{\eqns}[2]{Eqs.(\ref{#1}-\ref{#2})}
\newcommand{\mc}{Monte Carlo}
\newcommand{\intl}{\int\limits}
\newcommand{\intk}{\intl_K}
\newcommand{\intu}{\intl_0^1}
\newcommand{\dmux}{d\mu(x)}
\newcommand{\suml}{\sum\limits}
\newcommand{\sumk}{\suml_k}
\newcommand{\prol}{\prod\limits}
\newcommand{\promu}{\prol_{\mu}}
\newcommand{\prok}{\prol_k}
\newcommand{\umu}{^{\mu}}
\newcommand{\lmu}{_{\mu}}
\newcommand{\xp}[1]{\left\langle\;#1\;\right\rangle}
\newcommand{\intinf}{\intl_{-\infty}^{\infty}}
\newcommand{\mod}{\mbox{mod}\;}
\newcommand{\bino}[2] 
  {\left(\begin{tabular}{c} $#1$ \\ $#2$ \end{tabular}\right)}
\newcommand{\df}{{\mathcal D}f}
\newcommand{\si}{\sigma}
\newcommand{\dortho}{D^{\mbox{{\small ortho}}}}
\newcommand{\vecn}{\vec{n}}
\newcommand{\svn}{_{\vecn}}
\newcommand{\ctp}{\cos 2\pi}
\newcommand{\stp}{\sin 2\pi}
\newcommand{\rtt}{\sqrt{2}}
\newcommand{\dfour}{D^{\mbox{{\small Fourier}}}}
\newcommand{\real}{\mbox{Re}}
\newcommand{\w}{\omega}
\newcommand{\al}{\alpha}
\newcommand{\be}{\beta}
\newcommand{\order}[1]{{\mathcal O}\left(#1\right)}
\makeatletter

  \ifx\undefined\operator@font
    \let\operator@font=\rm
  \fi

  \def\Re{\mathop{\operator@font Re}\nolimits}
  \def\Im{\mathop{\operator@font Im}\nolimits}

\makeatother


\begin{flushright}
NIKHEF 96-017 
\end{flushright}
\nopagebreak

\begin{center}
\begin{Large}
  {\bf Multidimensional sampling for simulation and integration:\\
       \vspace{\baselineskip}
       measures, discrepancies, and quasi-random numbers}
\end{Large}\\
\vspace{\baselineskip}
{\bf F.~James${}^{1}$\\
     CERN, Geneva, Switzerland\\ 
     J.~Hoogland${}^{2}$\\
     NIKHEF-H, Amsterdam, the Netherlands\\
     R.~Kleiss${}^{3}$\\
     University of Nijmegen, Nijmegen, the Netherlands}\\
\footnotetext[1]{e-mail: F.James@cern.ch}
\footnotetext[2]{e-mail: t96@nikhef.nl,$\;$ 
                 research supported by the Stichting FOM}
\footnotetext[3]{e-mail: kleiss@sci.kun.nl}
\vspace{2\baselineskip}
{\bf Abstract}
\end{center}
This is basically a review of the field of Quasi-Monte Carlo 
intended for computational physicists and other potential users
of quasi-random numbers.  As such, much of the material is not new, 
but is presented here in a style hopefully more accessible to physicists
than the specialized mathematical literature. 
There are also some new results: 
On the practical side we give important empirical properties of large
quasi-random point sets, especially the exact quadratic discrepancies;
on the theoretical side, there is the exact distribution of
quadratic discrepancy for random point sets.
\\

\vspace{2cm}
\begin{center}
{\it Preprint accepted for publication in Computer Physics Communications}
\end{center}


\newpage


\section{Introduction}

In numerical integration, the magnitude of the 
integration error depends on two things: 

\begin{enumerate}
\item
The fluctuating behaviour(variance,variation) of the integrand. 
This behaviour may be improved
by so-called {\em variance-reduction techniques}, 
which we shall not discuss in this paper (see, for example,\cite{james80}). 
\item 
The quality of the point-set employed for the integration.
We shall see that the measure of quality will be the uniformity of the
point-set.
This paper deals with the  uniformity properties of point sets for numerical
integration: in particular,
we discuss various definitions of uniformity, 
their relation to typical integrands encountered,
some of their theoretical properties, 
and the actual performance of algorithms 
(quasi-random number generators) claimed to yield very uniform 
point sets. 
\end{enumerate}

The outline of this paper is as follows: 
in section 2, we discuss the traditional Monte
Carlo approach to numerical integration and relate the relatively slow  
convergence of the error, as  $1/\sqrt{N}$ (where $N$ denotes the size of the 
point set), to the inherent non-uniformity  of the truly random points, used
in classical Monte Carlo.
This is made explicit by the notion of {\em classical discrepancy\/} of the 
point set; we quote several theorems relating this discrepancy to the 
integration error. From these it is seen, that low-discrepancy point sets lead
to small errors. 
In section 3, we describe a number of algorithms that have 
been proposed for constructing {\em quasi-random sequences\/}, that lead
to point sets with a discrepancy smaller than that expected for truly
random point sets.
In general, for these algorithms, there exist discrepancy bounds, that are 
valid only in the limit $N\to\infty$. (Their observed behaviour for finite $N$
is reported in section 6.)
In section 4, we point out that the definition of classical
discrepancy is intimately connected with the underlying, implicit,
model for the type of integrand attacked; this {\em function class\/} is
governed by the Wiener measure, which may or may not be appropriate,
depending on the context of the integration problem.
We describe a number of alternative choices of function class, each
leading to its own definition of discrepancy.
In section 5, we argue that the merit of a `low-discrepancy` sequence
can only be judged in relation to a truly random point set. To this
end, one has to know the probability distribution of a given
discrepancy, viewed as a stochastic  variable through its dependence
of random points. We derive several results in this direction.
Finally, in section 6, we compute the classical discrepancy for a number of
quasi-random sequences for moderately large point sets ($N$ up to
150000), and moderately large dimension (up to 20).   


\section{Integration and discrepancy}
\subsection{Numerical integration}

The problem we shall be considering is the estimation of the 
multidimensional integral

\bq
J = \intk\;\dmux\;f(x)\quad,
\eq
where $x = x\umu = (x^1,x^2,\ldots,x^s)$ denotes a point in 
the $s$-dimensional integration region $K$.
Throughout this paper, 
Greek indices like $\mu$ will denote individual coordinates.
The integrand is $f(x)$, and $\dmux$ denotes a measure on $K$, that is, 
we must have
\bq
\intk\;\dmux = 1\quad,
\eq
and $\dmux$ must be positive. We shall only consider measures that
allow a probabilistic interpretation, that is, there exists,
{\em in principle,\/} a computer algorithm for turning a sequence of
truly random numbers into a series of points $x$ such that their
probability density is given by $\dmux$:
\bq
\mbox{Prob}(x\in A) = \intl_A\;\dmux\quad,
\eq
for all small rectangular regions $A$.

We estimate $J$ using the unweighted sum:
\bq
S = {1\over N}\sumk f(x_k)\quad,
\eq
where the $x_k$ constitute a {\em point set,\/} that is, a
finite set of $N$ points $x_k$, $k=1,2,\ldots,N$ obtained in one or
another way. Note that Latin indices like $k$ run over the points in
the point set and the sum is always understood to run from 1 to $N$. 
The error made in using this estimate is clearly
\bq
\eta = S - J\quad.
\eq

The object of this paper is to find point sets $x_k$ which can be 
used to estimate the integrals of general classes of functions $f$ 
with small error $\eta $.

We restrict the choice of integration regions $K$ to
the unit hypercube $I_s=[0,1)^s$ since this is the most common 
and best understood region, but we should point out that the properties
of many integration methods may depend strongly on the shape
of the integration region.  

Finally, throughout this paper we shall adopt the real-number model
of computation, that is, we shall disregard all kinds of small corrections
related to the finite word size of our computers, or the fact that
the numbers we use cannot be truly irrational: we assume that our
computers have enough precision to approximate real numbers for
all practical purposes.

\subsection{Monte Carlo integration: the Central Limit Theorem}

In the \mc\ method the point set is taken to be a sample of 
random points, independent and identically distributed $(iid)$
with probability density  $\dmux$. 
The error $\eta$ is then also a random quantity, with some 
probability density $P(\eta)$. This is most easily studied in the form 
of a {\em moment-generating function G\/}: with $M=\sqrt{N}$, we can write
\bq
G(Mz) \equiv \xp{e^{Mz\eta}} = \xp{e^{-MzJ}\prok e^{zf(x_k)/M}}\quad,
\eq
where the brackets denote an average over all point sets 
consisting of $N$ points
chosen with the measure $\dmux$ as a probability density.
Since the points $x_k$ are random, they are independent, and we can easily
take the average:
\bq
\xp{e^{zf(x_k)/M}} = \suml_{n\ge0}{z^n\over n!M^n}J_n\quad\;,\quad\;
J_n \equiv \intk\;\dmux\;f(x)^n\quad.
\label{independentaverage}
\eq
For large enough $N$, the following approximation is justified:
\bqa
G(Mz) & = & \left[e^{-zJ/M}\suml_{n\ge0}{z^n\over n!M^n}J_n\right]^N\nl
& = & \left[1 + {z^2\over2N}(J_2-J^2)+{z^3\over6NM}(J_3-3J_2J+2J^3)
 +\cdots \right]^N\nl
& = & \exp\left({z^2\over2}(J_2-J^2)\right)(1+{\mathcal O}(1/M))\quad.
\eqa
Up to terms of order $1/M$, we may compute $P(\eta)$ by Fourier
transform:
\bqa
P(\eta) & = & {1\over2\pi}\intinf\;dz\;e^{-iz\eta}G(iz) 
\;\; \sim \;\; {1\over\sqrt{2\pi V/N}}
\exp\left(-N\eta^2\over2V\right)\quad,\nl
V & = & J_2-J^2\quad.
\eqa
This is the Central Limit Theorem: the error is normally distributed
around zero, with standard deviation $\sqrt{V/N}$; 
$V$ is called the {\em variance\/} of the integrand: it is positive
for every $f(x)$ that is not constant over $K$.
We note the following aspects of the above derivation.
\begin{itemize}
\item The \mc\ estimate is {\em unbiased\/}: the expectation value
      of the error is zero (this holds also for finite $N$).
\item The error distribution attains a simple form owing to the
      `large-$N$ approximation', that is, we neglect terms of order $1/N$.
\item The result for the expected square error is rigorous:              
      but its use as an estimate for the error made using {\em this
      particular point set\/} implies that we postulate this point set to
      be a `typical' member of the ensemble of point sets governed by
      the underlying measure $\dmux$.
\item The error estimate does not depend on the particularities of the
      point set actually used, since we have integrated over all such 
      point sets.
\item The error estimate {\em does\/} depend on the particularities
      of the integrand, namely its variance, which is a measure of the
      amount by which it fluctuates over $K$.
\item Any quadratically integrable function can be integrated by \mc\,
      but the convergence of the error to zero is (only) as $1/\sqrt{N}$,
      and additional smoothness properties, such as continuity or
      differentiability only lead to smaller error inasmuch as they
      lead to a smaller $V$.
\end{itemize}

\subsection{Classical discrepancy}
In the previous section we have established the $1/\sqrt{N}$ convergence
of classical \mc, but it is known that, in one dimension,
even the unweighted quadrature method known as 
the trapezoid rule, where the points are
equidistantly distributed, gives a $1/N^2$ convergence.
Clearly, the irregularities in the random, rather than equidistant, 
point set play a r\^{o}le in the behaviour of the error. 
We now introduce the {\em classical discrepancy\/}, which will 
allow us to quantify such irregularities, not only in one dimension,
but also in higher dimensions where the trapezoid rule 
is outperformed by Monte Carlo.

Let $K$ be the $s$-dimensional hypercube $I_s$, and $y$ be a point in $K$.
We also assume that $\dmux$ is the usual Cartesian measure on $K$:
random points under this $\dmux$ have a uniform distribution
over $K$. We define the following counting function:
\bq
\chi(y;x) = \promu\theta(y\umu-x\umu)\quad,
\eq
which simply checks if the point $x$ is `below $y$', 
{\it i.e.\/} inside the hyper-rectangle
defined by the origin and $y$. The {\em local discrepancy} at $y$, for
the point set $x_k$, is then
\bq
g(y) = {1\over N}\sumk\chi(y;x_k) - \promu y\umu\quad.
\label{wienerlocaldisc}
\eq
The function $g(y)$ counts the fraction of the point set is below $y$, 
and compares this with the volume of $K$ that is below $y$. 
{}From a slightly different point of view, $g(y)$ is nothing
but the integration error made in estimating the volume below $y$
using the given point set. Clearly,
the more uniformly distributed the point set, the smaller $g(y)$.
This suggests the notion {\em global discrepancy}, that must inform us
about the global deviation of $g(y)$ from zero. There exist
various such quantities. Let us define
\bq
D_m \equiv \intk\;dy\;g(y)^m\quad.
\eq
Then, useful measures of global discrepancy are $D_1$ (linear discrepancy),
$D_2$ (quadratic discrepancy), and the extreme, or Kolmogorov discrepancy:
\bq
D_{\infty} = \lim_{k\to\infty} \left(D_{2k}\right)^{1/2k} =
\sup_{y\in K}|g(y)|\quad.
\eq
This last discrepancy (called $D^{\star}$ in \cite{kuipers}), has been
the most widely studied, but $D_2$ appears to be easier to evaluate
for a given point set (as we shall see), and may be more relevant as well.

We shall derive some results on the discrepancies for random points later on,
but at this point we may already quote the various expectations
for truly random point sets:
\bq
\xp{D_1}  =  0 \;\;,\;\;
\xp{D_2}  =  {1\over N}\left({1\over2^s}-{1\over3^s}\right)\;\;,\;\;
\xp{D_{\infty}} \stackrel{s=1}{=}  \sqrt{\pi\over2N}\log{2}\quad.
\eq

\subsection{Results for integration errors}
It should be intuitively clear that the size of the discrepancy of a point
set must relate to how good that point set is for numerical integration.
This has indeed been studied extensively; see, for example 
\cite{kuipers}  
from which we give some important results.

In the first place, if a point set has extreme discrepancy equal to $D$, 
every rectangular region (with edges parallel to the axes) of volume
$2^sD$ or larger must contain at least one point\footnote{The
factor $2^s$ is due to the restriction, in the definition of
discrepancy, to hyper-rectangles touching the origin.}: the
largest rectangular `gap' (with no points) has size $2^sD_{\infty}$.
The discrepancy is seen to measure the `resolution' of the point set. 
Secondly, let us define the `extremum metric' $d$ on $K$ by
\bq
d(x,y) = \sup_{\mu=1,2,\ldots,s}|x\umu-y\umu|\quad,
\eq
for any two points $x$ and $y$ in $K$. The {\em modulus of continuity\/}
$M$ of the integrand is then the following function:
\bq
M(z) = \sup_{x,y\in K} |f(x)-f(y)|\quad\;,\quad\;
\mbox{for $d(x,y)\le z$}\quad.
\eq
That is, the modulus of
continuity tells us what is the maximum jump in the value of $f$, if
we make a step of maximal size $z$. Then, when using the point set
to estimate $J$ by the quantity $S$, the error $\eta$ obeys
\bq
|\eta| \le 4M(D_{\infty}^{1/s})\quad;
\eq
the appealing result that the error depends on the maximal jump in
value, for steps of half the size of the edge of the hyper-cubic `gap' 
that corresponds to the resolution of the point set.

Although this result is only valid for continuous functions, for
which $M(z)\to0$ as $z\to 0$, we may always approximate a discontinuous
function by a continuous one: but then, $M(z)$ will no longer vanish with
$z$, and we have a finite error bound even for zero discrepancy.

A large number of other, similar results are given in \cite{kuipers}, for
instance the Koksma-Hlawka bound which relates the error to the
{\em variation\/} of the integrand, rather than to its variance: but,
since for discontinuous functions the variation is in many cases infinite
even when the variance is finite, such a bound may be of limited value;
moreover, quantities like the modulus of continuity or the variation
are in practice very hard to compute exactly.
Nonetheless, these results teach us that the lower the discrepancy, 
the smaller we can expect the integration error to be.

\subsection{Quasi-random Sampling for Simulation}

Formally at least, most simulations are in fact equivalent to
numerical integrations, since one is usually looking for the
expectations of particular variables or of distributions of variables
and these expectations are simply integrals over some subspaces
(histogram bins) of the full sampling space.
However, in practice, additional considerations may be important.
In simulations of particle physics experiments for example,
the dimensionality as defined by random numbers per event
tends to be extremely high and often is not even the same 
from event to event.  Clearly the results of this paper are not 
then immediately applicable, but one somehow suspects that there must
nonetheless be a way to take advantage of the improved uniformity
of quasi-random points in such simulations.

A promising way to approach this problem is to decompose the full
simulation of an event into manageable parts, some of which may
then be of reasonably low fixed dimension.  For example, the
physical event must be generated, and independently some of the 
particles may interact or decay, and independently they may produce
detector hits, showers, etc.  It may be advantageous to use quasi-random
points in the simulation of one or more of these sub-processes. 

Another difficulty in simulation is that the detector boundaries or
decision boundaries tend to be discontinuous, which formally makes
the problem equivalent to the numerical integration of a 
multidimensional discontinuous function, and the variation of
such a function is generally unbounded.  
Here again, practical experience indicates that the situation is
not as grim as all that, and that such integrals do in fact converge
with a considerably smaller error using quasi-random points 
\cite{schlier}.

\section{Low-discrepancy point sets}
\subsection{Finite sets and infinite sequences}
Much effort has been devoted to construct point sets that have
lower discrepancy than those consisting of random points. 
We must distinguish between two cases. 
On the one hand, one may consider {\em finite\/} point sets, 
where $N$ is fixed. 
In principle, for each value of N in the $s$-dimensional hypercube,
there exists a point set with the lowest discrepancy of all point sets. 
That discrepancy must be larger than the so-called Roth bound
\cite{kuipers}:
\bq
D_{\infty} \ge C_s{(\log N)^{(s-1)/2}\over N}\quad,
\label{rothbound}
\eq
where $C_s$ depends only on $s$. 
For large $N$ and reasonable $s$, this lower bound is much lower than 
the expected discrepancy for random point sets
but, except in one dimension, it is not known how to construct such 
an `optimal' set, and it is not even known if it is possible to attain
this bound in general. 
The one-dimensional exception is the equidistant point set:
\bq
x_k = {2k-1\over2N}\quad,\quad k=1,2,\ldots,N\quad,
\label{equidistant}
\eq
or any rearrangement of this set. 
In higher dimension, Cartesian products of this sequence,
the so-called {\em hyper-cubic lattice,\/} are {\em not\/}
optimal. For, let us enumerate a hyper-cubic sequence by
\bq
x\umu_{(k^1,k^2,\ldots,k^s)} = {2k\umu - 1\over2M}\quad\;,\quad\;
k\umu=1,2,\ldots,M\quad\;,\quad\;M^s=N\quad.
\eq
For this point set, we indeed have $D_1=0$, but
\bq
D_2 = {1\over3^s}\left[1+
\left(1+{1\over2M^2}\right)^s-2\left(1+{1\over8M^2}\right)^s\right]\quad,
\eq
which only goes as $N^{-2/s}$ for large $N$. This discrepancy is
larger than that expected for random points if $s>2$: and this is
the underlying reason why, in higher dimension, trapezoid rules
generally perform less well than \mc\ for the same number of points.

It is in principle possible to use numerical minimization techniques 
to find approximately optimal point sets, 
but the computational complexity of this problem is so enormous that 
it can be attempted only for very small values of both $N$ and $s$.

On the other hand, we usually do not wish to restrict ourselves in advance
to a particular value of $N$, but rather prefer to find a formula
which yields a long sequence of points for which the discrepancy is low
for all $N$, perhaps even approaching an optimal value asymptotically.
This allows us to increase the size of our point set at will until
the desired convergence is attained.
We shall, therefore, mainly concentrate on sequences of indefinite length.
Their discrepancy for any given $N$ will usually be larger than 
that of an optimal point set with the same $N$; 
but, `on the average', we may hope to remain close to the Roth bound.

\subsection{Richtmyer sequences}
Probably the first quasi-random point generator was that attributed to 
Richtmyer and defined by the formula for the $\mu$th coordinate 
of the $k$th point:
\bq
   x_k^\mu = [kS_\mu] \mod 1 \;\;,
\eq
where the $S_\mu$ are constants which should be irrational numbers in order
for the period to be infinite.  Since neither irrational numbers nor
infinite periods exist on real computers, the original suggestion was
to take $S_\mu$ equal to the square root of the $\mu$th prime number,
and this is the usual choice for these constants.

\begin{figure}
\begin{center}
\mbox{\hspace{-0.5cm}
  \input{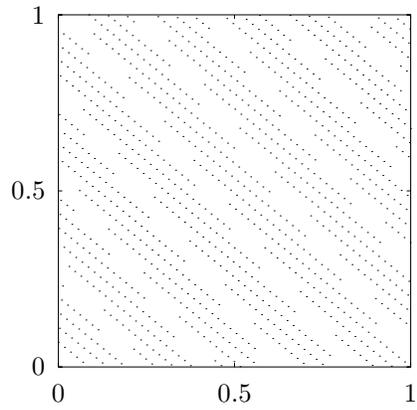}\hspace{-1cm}
  \input{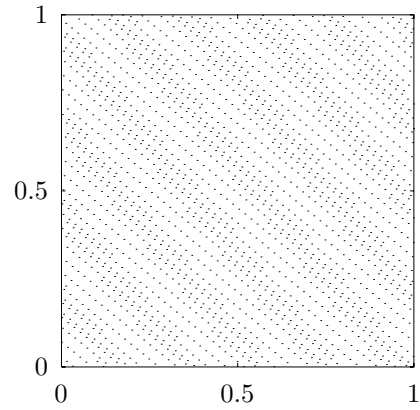}}
\mbox{\vspace{0.5cm}}
\mbox{\hspace{-0.5cm}
  \input{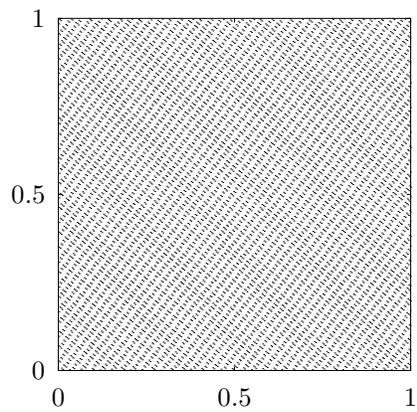}\hspace{-1cm}
  \input{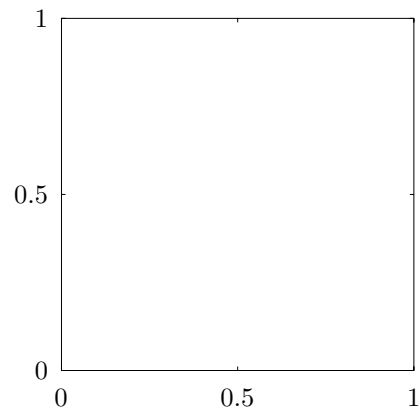}}
\end{center}
\caption[]{The distribution of Richtmyer points in two 
dimensions, showing the evolution as $N$ increases.}
\label{richtpts}
\end{figure}

Figure~\ref{richtpts} shows how the distribution of
Richtmyer points in two dimensions evolves as N increases.
A very regular band structure develops. The bands then become wider
until they completely fill the empty space. Then they start to overlap and
begin to form other bands in other directions which also widen, 
and the process continues with each successive band structure becoming
narrower and more complex.  
This kind of behaviour is a general property of quasi-random sequences,
even those described below which are based on a very different algorithm.

This simple quasi-random generator does not seem to have 
received much attention, even though it is the only one we know based 
on a linear congruential algorithm (others are based on radical-inverse
transformations), and as we shall see below, its
discrepancy is surprisingly good. 
{}From a practical point of view, some potential users have 
probably been discouraged by plots of points in two-dimensional 
projections which show some long-lasting band structure in some 
dimensions,
due of course to an unfortunate combination of constants $S_\mu$. 
In this paper we use always the square roots of the first prime 
numbers, but we should point out that it may be possible to
find other values of $S_\mu$ which assure a long period and
minimize the band structure or even minimize directly the discrepancy.
This is of course a very compute-intensive calculation.


\subsection{Van der Corput (or Halton) sequences}
An interesting low-discrepancy sequence can be found as follows
\cite{corput}.
Let us start, in one dimension, by choosing a base, an integer $b$. 
Any integer $n$ can then be written in base $b$:
\bq
n = n_0 + n_1b + n_2b^2 + n_3b^3 + \cdots\quad.
\eq
The {\em radical-inverse transform\/} (to base $b$) is defined by
\bq
\phi_b(n) = n_0b^{-1} + n_1b^{-2} + n_2b^{-3} + n_3b^{-4} + \cdots\quad.
\eq
The {\em van der Corput sequence\/} to base $b$ is then simply the following:
\bq
x_k = \phi_b(k)\quad\;,\quad\;k=1,2,\ldots\quad.
\label{corputonedim}
\eq
Note that, as $k$ increases, the digit $n_0$ changes most rapidly,
the digit $n_1$ the next rapidly, and so on: so that the leading
digit in $x_k$ changes the most rapidly, the next-to-leading one
less rapidly, and so on. It is therefore clear that this sequence
manages to fill the unit interval $(0,1)$ quite efficiently. Another
insight may be obtained from the fact that, when $N=b^m$, 
for integer $m$,
the set $x_1,x_2,\ldots,x_N$ will precisely be equidistant, of the
type of \eqn{equidistant}, and for that $N$ the discrepancy will
be optimal. The discrepancy is therefore optimal 
whenever $N$ is a power of $b$.
For other values of $N$, it cannot stray away too much from the
optimal value, but as $m$ increases it can go further and further out
before returning: it is therefore no surprise that the discrepancy
has a low upper bound:
\bq
D_{\infty} \le C_b{\log N\over N}\;\;,\;\;
C_b = \left\{\begin{tabular}{l l}
 ${b^2\over4(b+1)\log b}$ & when $b$ is even \\
 $\vphantom{X}$ & \\
 ${b-1\over4\log b}$ & when $b$ is odd
 \end{tabular} \right.
\label{corputdisconedim}
\eq
 
\begin{figure}
\begin{center}
\mbox{\hspace{-0.5cm}
  \input{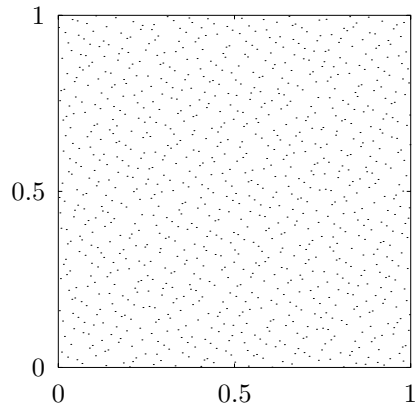}\hspace{-1cm}
  \input{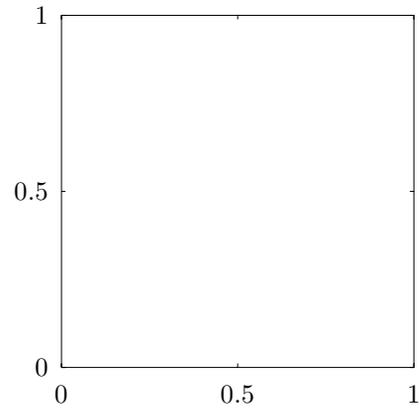}}  
\mbox{\vspace{0.5cm}}
\mbox{\hspace{-0.5cm}
  \input{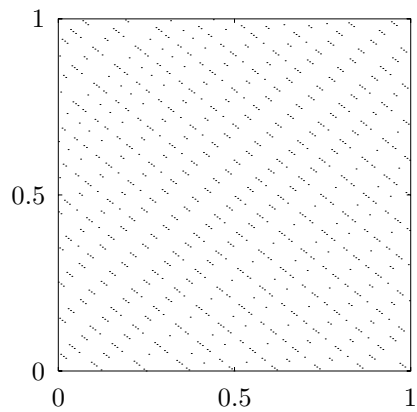}\hspace{-1cm}
  \input{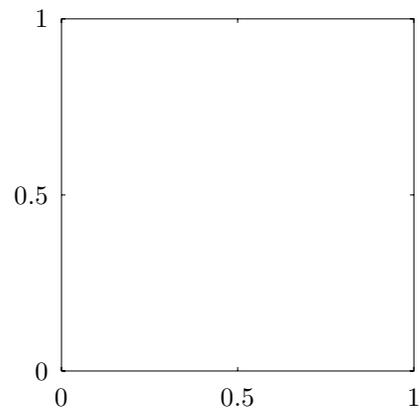}}  
\end{center}
\caption[.]{The first van der Corput points with bases 2 and 3 (top)
 and bases 17 and 19 (bottom).}
\label{corputpts}
\end{figure}

A generalization of \eqn{corputonedim} to higher dimensions is 
obvious: one simply chooses several bases $b_1,b_2,\ldots,b_s$, and 
\bq
x_k\umu = \phi_{b\lmu}(k)\quad\;,\quad\;
k=1,2,\ldots\quad\;,\quad\;\mu=1,2,\cdots,s\quad.
\label{corputmoredim}
\eq
In figure~\ref{corputpts}, 
we show the distribution, in two dimensions, of the
first 1000 points obtained with $b_1=2$, $b_2=3$. 
This distribution is, even visually, smoother and more uniform than
a typical set of 1000 random points (see figure~\ref{ranluxpts}).
Clearly we will get into trouble if any two bases have common 
factors, and indeed the most common choice (and our choice here)
is to take as bases the first $s$ prime numbers.
However, for dimensions 7 and 8 for instance, we then need  
bases 17 and 19, for which the plot is given in the bottom half
of figure~\ref{corputpts}: a lot of structure
is still visible, and the square is not filled as uniformly.


This can also be understood easily: the `recurrence' of the
discrepancy to a low value, discussed for $s=1$, could be encountered
here only if $N$ is an exact power of each of the bases, and since these
are all relatively prime, this cannot happen. Numbers that are `close'
to exact powers of all the bases (in the sense mentioned above) are
guaranteed to be far apart, and hence the discrepancy can become
considerable. This is in fact the content of the multidimensional
equivalent of \eqn{corputdisconedim}:
\bq
D_{\infty} \le {s\over N} +
{(\log N)^s\over N}\promu\left(
{b\lmu-1\over2\log b\lmu} + {b\lmu+1\over2\log N}\right)\quad.
\eq
As a function of $s$, therefore, the constant factor
grows faster than an exponential of $s$, and
in high dimensions the superiority of the van der Corput sequence
over a purely random one will appear only for impractically large $N$.\\

\subsection{Sobol' and Niederreiter sequences}

Improved discrepancy in higher dimensions can be obtained
from the van der Corput sequences by
making use of additional functions $m\lmu(k)$, $\mu=1,2,\ldots,s$,
and defining a new sequence 
\bq
x_k\umu = \phi_b(m\lmu(k))\quad,
\eq
The functions
$m\lmu(k)$ are cleverly chosen, amongst other things in such a way that
when $N=b^m$, the numbers $m\lmu(1),m\lmu(2),\ldots,m\lmu(N)$ are
just a permutation of $(1,2,\ldots,N)$. In this way, one can attain
a constant factor ($C_s$ in \eqn{rothbound})
that in principle {\em decreases\/} super-exponentially with $s$.
Practical experience with these sequences is still limited, however.

We have investigated empirically the behaviour of two such sequences:
the Sobol' sequence, described in \cite{sobol},
and the Niederreiter base-two sequence, described in \cite{niederreiter}.
The general Niederreiter sequences form a large class of which both the
above are special cases.

\section{Problem classes and discrepancies}
\subsection{A conceptual problem for special point sets}
We have seen how, in classical \mc, the error estimate is based
on the assumption that the point set is one out of an ensemble of
such, essentially equivalent, point sets: we then average over
this ensemble. We have also seen that the lower a point set's
discrepancy, the smaller the error could be. But when we try to
apply this statement to the integral of a particular function,
we immediately run into trouble. Namely, low-discrepancy point sets
are not `typical', they are special. In particular, the points are
correlated in a possibly very complicated way, so that the simple
treatment of \eqn{independentaverage} cannot be valid.
One way out would be to define, in some manner, an ensemble of
point sets with a given discrepancy.
This approach is followed in \cite{hooglandkleiss}.

Another option, which we shall pursue in the following, is to respect
the specialness of the point set, and instead consider the integrand
$f(x)$ to be a `typical' member in some set: this we shall call the
{\em problem class}. An example of such a class could be that of
all polynomials of some degree, but we shall encounter other, more
relevant, cases. Then, we may hope, instead of averaging over the
ensemble of point sets to end up with an error estimate depending
on the integrand, to integrate (or average) over the problem class:
in that case, we will end up with an error estimate that is independent
of the particular integrand, but related to some property of the
point set, for instance its discrepancy as discussed above.
It should be clear that different problem classes will lead to
different measures of irregularity of the point set: we shall
call such a measure the {\em induced discrepancy\/} corresponding 
to the problem class.

To pursue this line of reasoning, we must therefore first consider
what constitutes a problem class, and this we shall do next.

\subsection{Green's functions and connected Green's functions}
It is our aim to determine the integration error by averaging
over all integrands in a given problem class.
Strictly speaking, we then have to define a 
measure $\df$ on this problem class, which defines the
probability density for members of the class. In fact, we can settle
for less: the kind of information we need about the problem class 
is the set of all {\em Green's functions\/} in the problem class:
\bqa
g_n(x_1,x_2,\ldots,x_n) & \equiv &
\xp{f(x_1)f(x_2)\cdots f(x_n)}_f\nl
& \equiv & \int\;\df\;f(x_1)f(x_2)\cdots f(x_n)\quad,
\eqa
where this time the brackets denote an average over the problem class:
it is the `typical' value expected for the product of integrand values
at the points $x_1,x_2,\ldots,x_n$. It is, of course, symmetric in
all its arguments.
Knowledge of the Green's functions will be sufficient for our purposes, 
even if we cannot (as in the case of the Wiener measure) write down
the form of a particular integrand in the problem class in any simple
manner. The kind of problem class we consider, and therefore the Green's
functions, must of course be determined by the nature of our integration
problem. The restriction to continuous (or even smoother)
functions, popular in the mathematical literature, is not so 
appropriate in particle physics phenomenology, where experimental
cuts usually imply discontinuous steps in complicated shapes.

Even more relevant for our purposes than the Green's functions themselves
are the {\em connected Green's functions\/}
$c_n(x_1,x_2,\ldots,x_n)$, which form the irreducible building blocks
of the Green's functions. We have
\bqa
g_0() & = & 1\quad,\nl
g_1(x_1) & = & c_1(x_1)\quad,\nl
g_2(x_1,x_2) & = & c_1(x_1)c_1(x_2) + c_2(x_1,x_2)\quad,\nl
g_3(x_1,x_2,x_3) & = & c_1(x_1)c_1(x_2)c_1(x_3) 
 + c_1(x_1)c_2(x_2,x_3) + c_1(x_2)c_2(x_3,x_1)\nl
 & & +\;c_1(x_3)c_2(x_3,x_1) + c_3(x_1,x_2,x_3)\quad,
\eqa
and so on; the following recursive definition holds:
\bqa
\lefteqn{g_n(x_1,x_2,\ldots,x_n) =}\nl
& & \suml_{\mathcal P}
\suml_{m=1}^{n}\;{c_m(x_1,z_2,z_3,\ldots,z_m)\over(m-1)!}
{g_{n-m}(z_{m+1},z_{m+2},\ldots,z_n)\over(n-m)!}\quad,
\eqa
where $(z_2,z_3,\ldots,z_n)$ is a permutation ${\mathcal P}$
of $(x_2,x_3,\ldots,x_n)$: 
the first sum runs over all $(n-1)!$ such permutations. 
The connected Green's functions are, of course, also totally symmetric.
The nice thing about connected Green's functions is that in many cases
there is only a finite number of non-vanishing ones; for instance,
for a Gaussian problem class such as the Wiener sheet measure, only
$c_2$ is nonzero: in that case $g_n$ is zero if $n$ is odd, and
$g_{2n}$ consists of $(2n)!/2^nn!$ terms, 
each a product of $n$ factors $c_2$.

\subsection{Induced discrepancy}
We now proceed to derive how the set of Green's functions induces
a quantity with the general properties of a discrepancy. 
If the integrand is chosen `at random' 
from the problem class, the integration error $\eta$ will again be 
random even for a fixed point set, and we can derive its expectation 
value and moments. Denoting averages over the problem class by 
brackets we have
\bqa
\xp{\eta^m}_f & = & \xp{(S-J)^m}_f
 =  \suml_{p=0}^m\bino{m}{p}(-1)^{m-p}M_{m,p}\quad,\nl
M_{m,p} & = & {1\over N^p}\suml_{k_1,k_2,\ldots,k_p}
  \intk d\mu(z_{p+1})\;\intk d\mu(z_{p+2})\cdots\nl
& & \cdots\intk d\mu(z_m)\;
  g_m(x_{k_1},x_{k_2},\ldots,x_{k_p},z_{p+1},z_{p+2},\ldots,z_{m})\quad.
\eqa
Now, we express the $g_m$ in the connected Green's functions. 
The combinatorial factors are quite involved, but the result can be
summarized as follows. The moment $\xp{\eta^m}$ consists of a sum
of all powers of all possible combinations of $c$'s, with each $c$
having as arguments all possible combinations of summed-over point set
members and integrated-over points in $K$. Let us define
\bqa
a_{n,k} & = & {1\over N^k}\;\suml_{k_1,k_2,\ldots,k_k}\;
\intk d\mu(z_{k+1})\;\intk d\mu(z_{k+2})\cdots\nl
& & \cdots\intk d\mu(z_n)\;
c_n(x_{k_1},x_{k_2},\ldots,x_{k_k},z_{k+1},z_{k+2},\ldots,z_{n})\quad.
\eqa
The various $a_{n,k}$ with the same $n$ will
always occur in the combination
\bq
d_n = \suml_{k=0}^n{(-1)^{n-k}a_{n,k}\over k!(n-k)!}\quad,
\eq
and the $m^{\mbox{th}}$ moment of $\eta$ is given by
\bq
\xp{\eta^m}_f = m!\suml_{\{\nu_n\}} {d_1^{\nu_1}\over\nu_1!}
{d_2^{\nu_2}\over\nu_2!} {d_3^{\nu_3}\over\nu_3!}\cdots\quad,
\eq
where the sets of integers $\{\nu_n\}$ are restricted by 
$\nu_n\ge0$, and 
$\nu_1+2\nu_2+3\nu_3+\cdots=m$. Note that, whenever $c_n$ is a constant,
$d_n$ is zero.

We may now introduce the moment-generating function for $\eta$:
\bq
\xp{e^{z\eta}}_f = \suml_{m\ge0}{z^m\over m!}\xp{\eta^m}_f
 = \exp\left(\suml_{m\ge1}z^md_m\right)\quad.
\eq
In many cases, we may be able to construct the actual probability
distribution $P(\eta)$ for $\eta$ by inverse Mellin transform. We are
assured, in fact, that this transform always exists -- after all,
if the integrands have {\em some\/} probability distribution over the
problem class, then the integration error also has {\em some\/}
distribution.

More important is the realization that, if the problem class is
Gaussian in the sense that the only nonzero connected Green's function
is $c_2$, then $P(\eta)$ is necessarily a normal distribution, and
we only have to know $\xp{\eta^2}$ to know all confidence levels.
In addition, note that whenever $d_1$ is nonzero, the integration
will be biased. If, in addition to $d_2$, also $d_4$ is nonzero, it
had better be negative -- but, if it is positive, the problem class
itself is not well-defined.\\

What, now, is the use of all this for practical error estimates?
This lies in the following {\em factorization property:\/} suppose that
we can find a function $h(x;y)$, with $x$ in $K$, and $y$ being
{\em some\/} parameter in {\em some\/} space $L$, such that
the connected Green's function can be expressed in terms of $h$ as
follows:
\bq
c_n(x_1,x_2,\ldots,x_n) = \intl_L\;dy\;
h(x_1;y) h(x_2;y) \cdots h(x_n;y)\quad:
\eq
then, we can easily derive that $d_n$ has the simple form
\bqa
d_n & = & {1\over n!} \intl_L\;dy\;H(y)^n\quad,\nl
H(y) & = & {1\over N}\suml_k h(x_k;y) - \intk\;\dmux h(x;y)\quad.
\eqa
The function $H(y)$ is the induced local discrepancy, and $d_n$ is
a measure of induced global discrepancy. $H(y)$ has precisely the form
of \eqn{wienerlocaldisc}: it measures how well the function $h(x;y)$
is integrated by the point set, for a given value of $y$: but,
note that $h(x;y)$ itself is not necessarily a member of the problem class.
As we shall see, many problem classes have the factorization property.
For such a problem class, we then have the following strategy available:
\begin{enumerate}
\item determine the Green's functions;
\item determine the connected Green's function $c_2$;
\item find a parameter $y$ and a function $h(x,y)$ such that factorization
      holds;
\item for a given point set, determine $H(y)$ and $d_2$;
\item we have then estimated the expected moment $\xp{\eta^2}$ for
      the integration error if we use this point set on an integrand
      picked at random from the problem set.
\end{enumerate}

\subsection{Orthonormal-function measures}
We shall now discuss how to apply the above formalism for a special
kind of problem class. As usual, we start with the one-dimensional case.
Let us take $K=[0,1]$, and $\dmux = dx$, the usual uniform measure.
Furthermore, we define a set of functions $u_n(x)$, $n=0,1,2,3,\ldots$,
that are {\em orthonormal\/}:
\bq
\intu\;dx\;u_m(x)u_n(x) = \delta_{m,n}\quad\;,\quad\;
u_0(x) = 1.
\eq
As an example, we may take the $u_m$ to be orthonormal polynomials, and we
shall presently discuss some other examples. We then define our
problem class to consist of all functions that can be written as a
linear combination of these orthonormal ones:
\bq
f(x) = \suml_{n\ge0} v_n u_n(x)\quad,
\eq
so that the coefficients $v_n$ completely determine $f(x)$.
If our orthonormal set is even {\em complete,\/} that is, if
\bq
\lim_{N\to\infty}\suml_{n=1}^N u_n(x_1)u_n(x_2) = \delta(x_1-x_2)\quad,
\eq
we can actually approximate any $f(x)$ to an arbitrary accuracy by
members of the problem class.

A measure on the problem class can now be specified by giving the
probability density of each individual coefficient $v_n$. For simplicity,
we take these distributions to be normal, and we can write
\bq
\df = \prol_{n\ge0} dv_n{1\over\sqrt{2\pi\si_n^2}}
\exp\left(-{v_n^2\over2\si_n^2}\right)\quad,
\eq
where $\si_n$ is the width, which may be different for different $v_n$.
Now, the only nonzero connected Green's function is $c_2$, and
we always have the factorization property:
\bqa
c_2(x_1,x_2) & = & \suml_{n\ge0}\si_n^2 u_n(x_1)u_n(x_2)\quad\;
 =\quad\;\intu dy\;h(x_1;y) h(x_2;y)\quad,\nl
h(x;y) & = & \suml_{n\ge0}\si_n u_n(x) u_n(y)\quad.
\eqa
Thus, we can immediately write down the induced discrepancy:
\bq
\dortho_2 = {1\over N^2}\hat{\suml_n}\si_n^2
\suml_{k,l} u_n(x_k)u_n(x_l)\quad,
\label{orthoinduceddis}
\eq
where the caret denotes a sum over all $n$ {\em except\/} $n=0$: this
is due to the fact that $u_0(x)=1$ is a constant, and obviously the
constant component of an integrand is always estimated with zero error
by any point set.\footnote{The quadratic discrepancy $\dortho_2$
is defined here as $2d_2$, such that it is the expectation value 
of $\xp{\eta^2}_f$.}

The extension to more dimensions is fairly straightforward, if we
replace the simple one-dimensional enumeration $n=0,1,2,\ldots$ by
a multidimensional one: 
\bq
n = \vecn = (n_1,n_2,\ldots,n_s)\quad\;,\quad n\umu=0,1,2,3,\ldots\quad.
\eq
Then, we simply put
\bq
u\svn(x) = \promu u_{n\umu}(x\umu)\quad,
\eq
and (for a Gaussian measure as above)  we again have
\bq
\dortho_2 = {1\over N^2}\hat{\suml\svn}\si\svn^2
\suml_{k,l} u\svn(x_k)u\svn(x_l)\quad,
\eq
where now the sum runs over all vectors $\vecn=(n_1,\ldots,n_s)$ except
the null vector $(0,\ldots,0)$.\\

In the following we shall discuss some explicit examples of these
problem classes, and the particular form of their induced discrepancies.
At this point, however, we may already note that,
if the orthonormal set is complete, and if $\si_n$ 
does not go to zero as $n\to\infty$, 
then $\dortho_2$ will be infinite, since it will have contributions
proportional to $\delta(0)$. 
In concrete terms, in such a problem class the typical integrands
show such wild behaviour that the average squared integration
error is infinite. In our explicit examples we shall see how
the behaviour of $\si_n$ relates to the smoothness properties 
(or lack thereof) of the functions in the problem class.

\subsection{Examples}
We shall now discuss some explicit examples of problem classes, and
their induced quadratic discrepancies. The first two examples are
not of the orthonormal-function type; the last two are.

\subsubsection{Wiener sheet measure: the Wo\'{z}niakowski Lemma}
One of the more popular problem class measures is the Wiener
measure. It is widely used in field theory and statistical physics,
and describes, basically, a Brownian motion. Although its precise
definition, especially in more dimensions, is subtle, we can
describe it in the following manner. Let us start with the case $s=1$, and
`discretize' the unit interval $[0,1)$ into many small subintervals
(not necessarily equal), to wit $[0,x_1)$, $[x_1,x_2)$, $[x_2,x_3)$, and
so on. The value of the function jump from $x_p$ to $x_{p+1}$ is
then assumed to be distributed normally around zero, with variance
given by the interval size (not its square!). In this way, the
Wiener measure becomes transitive, in that the distribution for
$f(x_{p+2})$, when we jump to $x_{p+2}$ directly from $x_p$, is the
same as that obtained when we first jump from $x_p$ to $x_{p+1}$, and then
from $x_{p+1}$ to $x_{p+2}$. Therefore, we may insert new interval endpoints
in-between the old ones at will, and so approach the continuum limit.
The {\em derivative\/} of the function will then have the Green's function
\bq
\xp{f'(x_1)f'(x_2)}_f = \delta(x_1-x_2)\quad.
\eq
This singular behaviour comes from the fact that the average value of the
jump in $f$ is proportional to the {\em square root\/} of the jump size, 
whereas for differentiable functions it would, asymptotically, be proportional
to the jump size itself. The Wiener measure is, therefore, completely
dominated by functions that are everywhere continuous but nowhere 
differentiable.

By integration, we can simply derive the Green's function for the
function values themselves:
\bq
\xp{f(x_1)f(x_2)}_f = \min(x_1,x_2)\quad.
\eq
In more dimensions, we have the {\em Wiener sheet measure,\/}
$K=I_s$, the r\^{o}le of the derivative $f'(x)$ is played
by the multiple derivative:
\bq
f_s(x)\equiv (d^s/dx_1dx_2\cdots dx_s)f(x)\quad,
\eq
and we have
\bqa
\xp{f_s(x_1)f_s(x_2)}_f & = & \promu \delta(x_1\umu-x_2\umu)\quad,\nl
\xp{f(x_1)f(x_2)}_f & = & \promu\min(x_1\umu,x_2\umu)\quad.
\eqa
In addition, we have $\xp{f(x)}=0$, so that the only nonzero
connected Green's function is
\bqa
c_2(x_1,x_2) & =  & \promu\min(x_1\umu,x_2\umu) \;\;=\;\;  \intk\;dy\;h(x_1;y) h(x_2;y)\quad,\nl
h(x;y) & = & \promu\theta(x\umu-y\umu)\quad,
\label{wienerinduceddis}
\eqa
where we have also indicated the factorization. Inspection
tells us that, in fact, $h(x;y)$ is equal to the function $\chi$
discussed for the classical discrepancy, only with the points $x$
and $y$ reflected: $x\umu\to1-x\umu$, $y\umu\to1-y\umu$. We therefore
arrive immediately at the so-called {\em Wo\'{z}niakowski lemma:\/}
{\it for integrands distributed according to the Wiener sheet measure,
the average squared integration error is equal to the classical
$D_2$ discrepancy, for the reflected point set.}
For random points, the expected value of 
the quadratic induced discrepancy is
\bq
D_2^{\mbox{{\small Wiener}}} = 
{1\over N}\left({1\over2^s}-{1\over3^s}\right)\quad,
\eq
as discussed above. The
reflection of the point set is in practice not very important
since any self-respecting point set is, to a high degree, more or
less invariant under reflection.

This result, published in \cite{woz}, was the first instance of
the relation between problem classes and discrepancies. It is seen
here to arise as a special case of a more generally applicable
induction of discrepancies. Since the Wiener measure as discussed
is Gaussian, we have also established that the error will, in fact,
be normally distributed over the problem class, which fact was not
noticed in \cite{woz}.

\subsubsection{Folded Wiener sheet measure}
As mentioned, integrands that are `typical' under the Wiener
measure may not be of the kind studied in particle physics
phenomenology: as an illustration, we present, in fig.6, a `typical'
integrand in one dimension. Although they are continuous, 
such integrands have no additional smoothness properties. 
 

\begin{figure}
\begin{center}
  \input{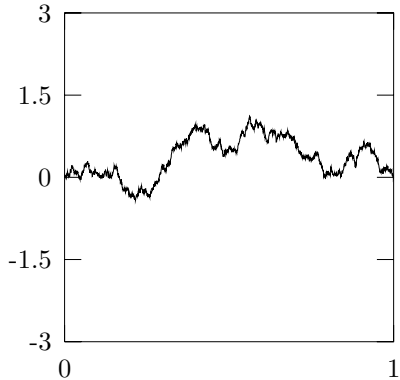}
\end{center}
\caption[.]{A typical integrand under the Wiener measure.}
\end{figure}


As an approximation to smoother functions, one may consider so-called
{\em folded Wiener sheet measures,\/} which are obtained by (repeated)
integration, so that the fractal zig-zag structure becomes hidden
in some higher derivative\cite{paskov}. A typical integrand under the $n$ times
folded Wiener sheet measure is defined, recursively, by
\bq
f^{(n)}(x) = \intl_0^{x_1}dx'_1\;\intl_0^{x_2}dx'_2\cdots
\intl_0^{x_s}dx'_s f^{(n-1)}(x')\quad,
\eq
where $f^{(0)}(x)$ is the integrand chosen from the Wiener measure.
For these folded measures, the induced discrepancy can therefore be
obtained trivially by repeated integration of $h$ in
\eqn{wienerinduceddis} over the variables $x$:
\bq
h^{(n)}(x;y) = \promu {(x\umu-y\umu)^n\over n!}\theta(x\umu-y\umu)\quad,
\eq
and, for random points, the expected quadratic induced discrepancy
in this case is given by
\bqa
D_2^{\mbox{{\small folded Wiener}}} & = & 
  {1\over N} \left(
               {1\over n!}
             \right)^{2s} 
             \left( 
               \promu {1\over(2n+1)(2n+2)}
               -
               \promu {1\over (n+1)^2(2n+3)} 
             \right)
\quad.\nl
\eqa
This discrepancy, and hence the expected integration error, decreases
extremely rapidly with increasing degree of folding. This can be understood
easily if we plot, say, the one times folded equivalent of fig.6, which
is given in fig.7. 
 

\begin{figure}
\begin{center}
  \input{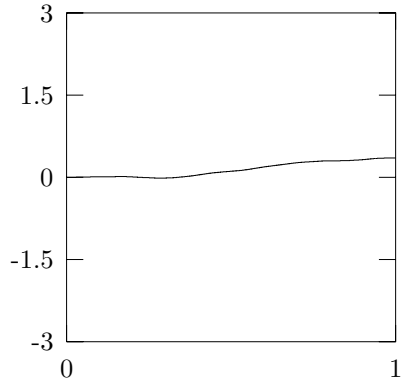}
\end{center}
\caption[.]{A typical integrand under the one time folded Wiener measure.}
\end{figure}


One sees that, even under only one folding, we
obtain functions that are extremely smooth, and it is not surprising that
such integrands can be treated with very small error. 
We conclude that, although the Wiener measure may be applicable in some
fields (the behaviour of the stock market, say), it is not really
a valid model of typical integrands in particle physics phenomenology.
 
\subsubsection{Fourier measure}
We shall now discuss another problem class measure, using orthonormal
functions as described above: again, first in one dimension.
Let us take
\bq
u_0(x) = 1\;\;,\;\;u_{2n}(x) = \rtt \ctp nx\;\;,\;\;u_{2n-1}(x) = \rtt \stp nx\;\;.
\label{fourierbase}
\eq
These functions are even a complete set: if we disregard the Gibb's 
phenomenon, which we may for purposes of integration, every practically
relevant function can be approximated arbitrarily closely by a combination
of $u_n$'s. In appendix B we present a short proof for completeness.
Also, in many physical cases the Gaussian distribution of the
magnitude of the several higher modes is quite reasonable.

We may immediately write down the form of the induced quadratic 
discrepancy, as given in \eqn{orthoinduceddis}. At this point, 
we may make another assumption. The two functions $u_{2n}$ and
$u_{2n-1}$ describe the mode with frequency $n$, with two phases
apart by $\pi/2$. These may be combined into a single mode, with
a single phase $x_0$:
\bqa
v_{2n-1}u_{2n-1}(x) + v_{2n}u_{2n}(x) & = & 
\rtt v_{2n-1}\stp nx + \rtt v_{2n}\ctp nx \nl
& \propto & \ctp n(x+x_0)\quad.
\eqa
Now, if we, reasonably, assume that $x_0$ is uniformly distributed between
$-\pi/2$ and $\pi/2$, we can infer that this necessitates
\bq
\si_{2n-1} = \si_{2n}\quad\;,\quad\;n=1,2,\ldots\quad,
\eq
and in the following we shall use this simplification. The discrepancy
then takes the form
\bqa
\dfour_2 & = & {2\over N^2}\suml_{n>0}\si_{2n}^2
\suml_{k,l}\ctp n(x_k-x_l)\nl
& = & {2\over N^2}\suml_{n>0}\si_{2n}^2
|\suml_k \exp(2i\pi nx_k)|^2\quad.
\label{fourierdis}
\eqa
Note that we have two equivalent expressions here, that are however of
different consequence in computer implementation. The first line
contains a double sum over the point set, and it can be evaluated
in time ${\mathcal O}(N^2)$, if we know a closed expression for
$\sum\si_{2n}^2\cos(2\pi nx)$. For instance, we may
assume $\si_{2n} = 1/n$, in which case we have
\bq
\suml_{n\ge1}{1\over n^2}\cos(2\pi n x) =
{\pi^2\over6}\left(\vphantom{{1\over1}}1-6x(1-x)\right)\quad.
\eq
On the other hand, we may use a single sum as in the second line
of \eqn{fourierdis}; but then, we have to sum over the modes explicitly.
Supposing that we only really need the modes up to $n=m$, the discrepancy
can then be computed in time ${\mathcal O}(mN)$.\\

The multidimensional extension is again rather trivial, if we assume
the orthonormal functions as a direct product of the one-dimensional ones, 
as discussed above. We may immediately write out the induced quadratic
discrepancy. With a little algebra, we may, in fact, arrive at
\bqa
\dfour_2 & = & {1\over N^2}\hat{\suml\svn}\tau\svn
|\suml_k\exp(2i\pi\vecn\!\cdot\!\vec{x}_k)|^2\quad,\nl
\tau_{(n^1,n^2,\ldots,n^s)} & = & \si^2_{(2|n^1|,2|n^2|,\ldots,2|n^s|)}\quad,
\eqa
where the sum over $\vecn$ now runs over {\em all\/} integer values
(both negative and positive) of the components, excluding the null
vector $(0,0,\ldots,0)$.\\
 

\begin{figure}
\begin{center}
  \input{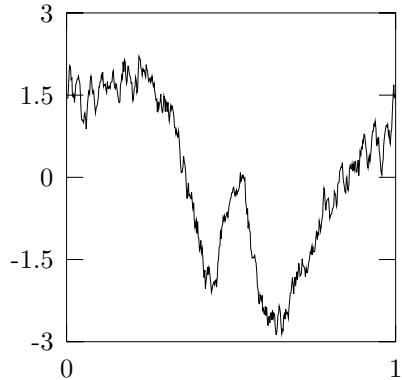}
\end{center}
\caption[.]{A typical integrand under the Fourier measure, with $\sigma_{2n-1}=\sigma_{2n}=1/n$.}
\end{figure}


We finish this section by pointing out a relation between our
results for the induced discrepancy and the exponential sums
discussed at length in \cite{kuipers,niederreiter} (for instance,
see \cite{kuipers}, chapter 2, section 2). There, the 
classical Kolmogorov discrepancy
is related to exactly the same kind of exponential, the only difference
being the absence of the factor $\tau\svn$. What does this imply?
Note that, in one dimension (for simplicity) we have
\bqa
\xp{\intu dx f(x)^2}_f & = & \suml_{n\ge0}\si_n^2\quad,\nl
\xp{\intu dx f'(x)^2}_f & = & 4\pi^2\suml_{n\ge0}n^2\si_{2n}^2\quad,\nl
\xp{\intu dx f''(x)^2}_f & = & 16\pi^4\suml_{n\ge0}n^4\si_{2n}^2\quad,
\eqa
so that the convergence of these sums informs us about the smoothness
of the typical integrands: the smoother, on average, the integrand, the faster 
the $\si_n$ has to approach zero. If $\si_n=1$, as implicit in
the treatment as given in \cite{kuipers}, the average integrand
is not even quadratically integrable! 
An example of such an integrand is given in 
fig.8. From the point of view of the 
present paper, choosing a slowly decreasing $\si\svn$ 
appears to lead to a certain kind
of overkill in the selection of good point sets.

\subsubsection{Walsh measure}
 

\begin{figure}
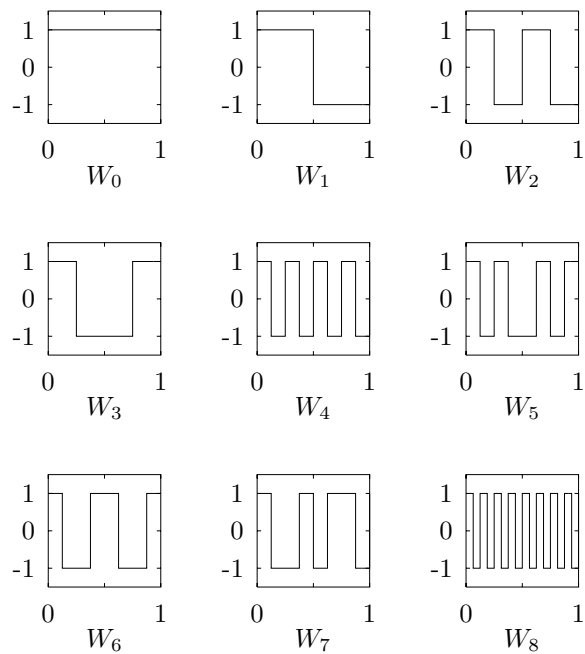

\begin{center}
\mbox{ 
  \input{figwalsh_0small}\hspace{-0.75cm}
  \input{figwalsh_1small}\hspace{-0.75cm}
  \input{figwalsh_2small}}
\mbox{ 
  \input{figwalsh_3small}\hspace{-0.75cm}
  \input{figwalsh_4small}\hspace{-0.75cm}
  \input{figwalsh_5small}}
\mbox{ 
  \input{figwalsh_6small}\hspace{-0.75cm}
  \input{figwalsh_7small}\hspace{-0.75cm}
  \input{figwalsh_8small}}
\end{center}
\caption[.]{The first nine Walsh functions, $W_0(x)$ to $W_8(x)$.}
\end{figure}


Now we shall discuss yet another complete set of orthonormal functions,
the so-called {\em Walsh\/} functions. Again, we start with $s=1$. 
We first introduce the Rademacher functions $\phi_n(x)$:
these are periodic with period one, and, for $0\le x<1$,
\bq
\phi_1(x) = \left\{\begin{tabular}{cc}
 1 & \mbox{if $x < 1/2$} \\ -1 & \mbox{if $x \ge 1/2$} 
\end{tabular}\right.\;\;,\;\;
\phi_n(x) = \phi_{n-1}(2x)\quad.
\eq
The Walsh functions $W_n(x)$ are now given as follows: let the
binary decomposition of $n$ be
\bq
n = n_1 + 2n_2 + 2^2n_3 + 2^3n_4 + \cdots\quad,
\eq
then $W_n(x)$ is
\bq
W_n(x) = \phi_1(x)^{n_1} \phi_2(x)^{n_2} \phi_3(x)^{n_3} \cdots\quad,
\eq
and we define $W_0(x)\equiv1$. As an example, we present the first
few Walsh functions in fig.9. 
The Walsh functions form a complete set, and we may
use them to construct a problem class, defined under the {\em Walsh
measure.\/} In fig.10, we show a typical integrand
under the Walsh measure, with $\si_n=1/n$. 
 

\begin{figure}
\begin{center}
  \input{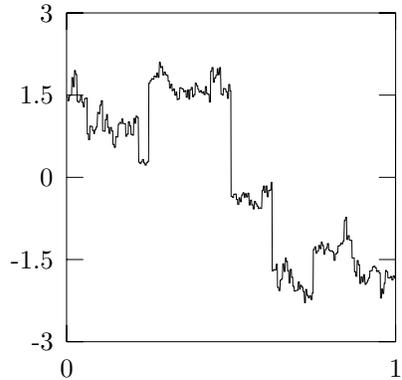}
\end{center}
\caption[.]{A typical integrand under the Walsh measure,
using $W_n$ with $n=0,1,\ldots,200$, and $\sigma_n=1/n$.}
\end{figure}


Some insight in the computational character of the Walsh functions
can be gained as follows. For any $x$ in the unit interval $[0,1)$,
let us define its binary notation by
\bq
x = x_12^{-1} + x_22^{-2} + x_32^{-3} + \cdots\quad;
\eq
it is, then, easy to see that
\bq
\phi_n(x) = (-1)^{x_n}\quad,
\eq
and therefore, the Walsh functions just measure certain parities: with $n$
decomposed as above, we have
\bq
W_n(x) = (-1)^{n_1x_1+n_2x_2+n_3x_3+\cdots}\quad.
\eq
Clearly, in a language with command of bitwise operations, such
an evaluation can easily be implemented. Note, moreover, that as long as
we restrict the range of $n$, our assumption of the real-number model of
computation is justified: if the least significant bit of $x$ has a value
of $2^{-p}$, then the finite word length of our computer is irrelevant
as long as $n<2^p$. 

The extension to the case of more dimensions is of course straightforward,
but we shall not discuss this further.

\section{Discrepancies for random point sets}
\subsection{Introduction: the need for comparison}
So far, we have studied how the choice of integration problem classes
relates to discrepancy-like properties of the integration point set.
Obviously, we should like to improve on classical \mc\ as much as we
can. This raises the question we shall study in the rest of this
paper: suppose we have a point set (or sequence) at our disposal, 
which we
want to use for integration. Suppose, moreover, that we have computed its
extreme, or another induced, discrepancy. To what extent is this
discrepancy better than what we could expect for random points?
To answer this question, we have to compute, for a given value of the
discrepancy, the probability that purely random points would give this
value (or less) for the discrepancy. That means, we have to know the
probability distribution of the discrepancy, which is now viewed as a
random quantity through its dependence on the set of random points.
Of course, we could restrict ourselves to the relatively simple 
question of the average discrepancy for random points, 
but it is much better
to know the confidence levels, and, moreover, we shall see that they can
actually be computed in many cases.
Another point is that we might use a source of (pseudo)random numbers
to produce a histogram of the discrepancy for a number of generated
point sets --- but in practice this is not feasible, since we would,
before generating such a histogram, have to ascertain that the
produced sequence is actually acceptably random: and how to do this
without comparing to the theoretical expectations?


\begin{figure}
\begin{center}
\mbox{\hspace{-0.5cm}
  \input{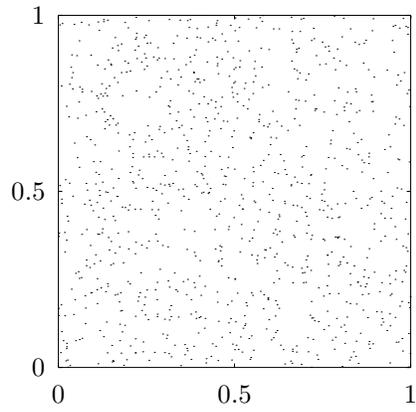}\hspace{-1cm}
  \input{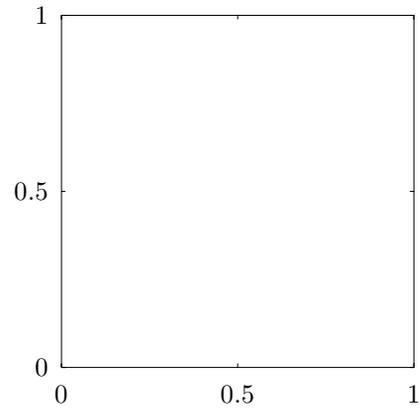}}
\mbox{\vspace{0.5cm}}
\mbox{\hspace{-0.5cm}
  \input{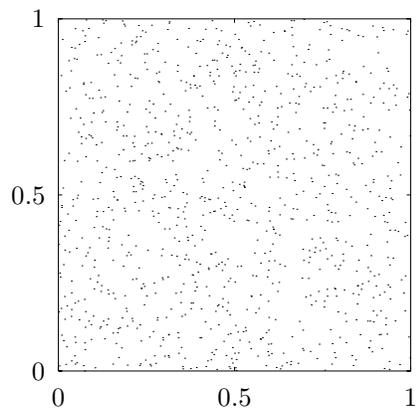}\hspace{-1cm}
  \input{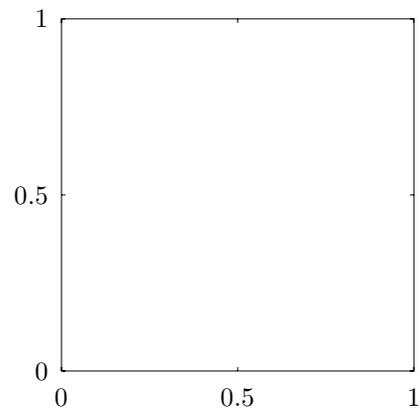}}
\end{center}
\caption[.]{Some points generated in two dimensions using a good
 pseudorandom generator (RANLUX, level 4 \cite{ranlux}).}
\label{ranluxpts}
\end{figure}


\subsection{Moment-generating function}
We shall study the moment-generating function of the linear and
quadratic induced discrepancies, that is: if for a set of $N$ truly
random points, the linear discrepancy equals $x$, we then try to
find the form of
\bq
G_1(z) = \xp{\exp(zx)}\quad,
\eq
and then, we can (numerically) find the probability density for $x$:
\bq
f_1(x) = {1\over2\pi}\intl_{-\infty}^{\infty}e^{-izx}G_1(iz)\quad;
\eq
and, similarly, we can find $G_2$ and $f_2$ for the quadratic discrepancy.
Although the linear discrepancy is not as important as the quadratic one, 
we shall discuss it first since it indicates how the limit of large
$N$ can simply be taken.

\subsubsection{Linear Discrepancy}
Let us assume that, for a given problem class, the discrepancy function
$h(x;y)$ is known. The linear discrepancy is then given by
\bq
D_1 =  {1\over N}\intl_L\;dy\;\sum_k\;\w_k(y)\;\;,\;\;
\w_k(y) =  h(x_k;y)-\intk\;\dmux\;h(x;y)\quad.
\eq
To get the moment-generating function, we have to evaluate
\bqa
\xp{D_1^p} & = & {1\over N^p}
\intl_Ldy_1\;\intl_Ldy_2\cdots\intl_Ldy_p
\xp{\suml_{k_{1,2,\ldots,p}}\w_{k_1}(y_1)\w_{k_2}(y_2)\cdots
\w_{k_p}(y_p)}\quad,\nl
\eqa
for all $p\ge0$. Let us consider the sum over the indices $k$ in detail.
All indices run from 1 to $N$, and some of the indices may be equal to
others. Those contributions where the indices fall into
$m$ distinct values (that is, some of the $k_i$ all have one value,
some other all have another value, a third group has another value,
and so on), enter with a combinatorial factor 
$N(N-1)(N-2)\cdots(N-m+1)$. It is clear that the largest combinatorial
factor is obtained if as many as possible of the indices $k$ are
different. Now, we have
\bq
\xp{w_k(y)} = 0\quad,
\eq
and the largest nonzero contribution is therefore the one coming
from the cases where the indices come in pairs. This also means that
$\xp{D_1^p}$ vanishes (to leading order in $N$) if $m$ is odd, and
\bq
\xp{D_1^{2m}} \sim {1\over N^{2m}}N^m{(2m)!\over2^mm!}
\left(\intl_Ldy_1dy_2\;\xp{w_k(y_1)w_k(y_2)}\right)^m\quad,
\eq
where we have also made the approximation 
\bq
N(N-1)(N-2)\cdots(N-m+1) \sim N^m\quad,
\eq
which is justified when $N\gg m^2$.  The moment-generating
function can now be written as
\bqa
\xp{e^{zD_1}} & = & \suml_{m\ge0}{z^{2m}\over(2m)!}\xp{D_1^{2m}}
 =  \exp\left({z^2\over2N}W^n\right)\quad,\nl
W & = & \intk\dmux\;\intl_Ldy_1dy_2\;h(x;y_1)h(x;y_2)\nl
 & & - \left(\intk\dmux\;\intl_Ldy\;h(x;y)\right)^2\quad,  
\eqa
and the probability density can be obtained by Fourier transform.
In the large-$N$ limit, the value of $D_1$ has a Gaussian distribution
with mean zero and width $\sqrt{W/N}$. Of course, all this is nothing but
the Central Limit Theorem, but we shall use the above combinatorial
arguments also in our study of the quadratic discrepancy.

The value of $W$ depends, of course, on the problem class.
The Wiener measure discussed above leads, in one dimension, to
\bqa
W & = & \intu\;dx dy_1dy_2\;\theta(y_1-x)\theta(y_2-x)
 - \left(\intu\;dx dy\;\theta(y-x)\right)^2\nl
& = & \intu\;dy_1dy_2\min(y_1,y_2) - \left(\intu\;dy\;y\right)^2\nl
& = & {1\over3} - {1\over4}\quad,
\eqa
and in higher dimension, we obviously have
\bq
W = \left({1\over3}\right)^s - \left({1\over4}\right)^s\quad.
\eq
In the case of problem classes defined using orthonormal functions
as above, we notice that
\bq
\intl_L\;dy\;h(x;y) = \suml_{n>0}\si_nu_n(x)\intl_Ldy\;u_n(y) = 0\quad.
\eq
and therefore $D_1$ is always strictly zero for such problem classes.

\subsubsection{Quadratic Discrepancy}
We now turn to the somewhat more complicated case of the quadratic
induced discrepancy. In order to avoid factors of $N$ cropping up
everywhere, we shall consider $ND_2$ rather than $D_2$ itself.
We have
\bq
D_2 = {1\over N^2}\intl_Ldy\;\suml_{k,l}\w_k(y)\w_l(y)\quad.
\eq
Its $m^{\mbox{{\small th}}}$ moment is given by
\bqa
\xp{N^mD_2^m} & = & {1\over N^{m}}\intl_Ldy_1\;\intl_Ldy_2\cdots\intl_Ldy_m
\suml_{k_{1,2,\ldots,m}}\suml_{l_{1,2,\ldots,m}}\nl
& & \xp{\w_{k_1}(y_1)\w_{l_1}(y_1)\w_{k_2}(y_2)\w_{l_2}(y_2)
\cdots\w_{k_m}(y_m)\w_{l_m}(y_m)}\quad.
\eqa
By the same combinatorial argument as before, in the large-$N$ limit
we only have to consider the contribution to the sum that consists
only of pairs of $\w$'s:
\bqa
\al(y_1,y_2) & \equiv & \xp{\w_k(y_1)\w_k(y_2)} \;\;=\;\; \be(y_1,y_2) - v(y_1)v(y_2)\quad,\nl
\be(y_1,y_2) & = & \intk dx\;h(x;y_1)h(x;y_2)\quad,\nl
v(y) & = & \intk dx\;h(x;y)\quad.
\eqa
At this point, a few remarks are in order. In the first place,
since every parameter $y_i$ occurs twice, the various pairs cannot
be considered independently: rather, we have to deal with objects of
the form
\bq
Z_n = \intl_L\;dy_1\;dy_2\cdots dy_n
\al(y_1,y_2)\al(y_2,y_3)\cdots\al(y_{n-1},y_n)\al(y_n,y_1)\quad.
\eq
Another consideration is that of factorization in more dimensions.
Whereas, in higher dimensions, the functions $h(x;y)$ themselves usually
factorize into a product of one-dimensional factors (as for the Wiener
and orthonormal problem classes), this no longer holds for the
functions $\al$: only the functions $\be$ and $v$ factorize.
We therefore decide to build up $\xp{D_2^m}$ from two distinct
objects:
\bqa
C_n & = & \intl_L\;dy_1\;dy_2\cdots dy_n
 \be(y_1,y_2)\be(y_2,y_3)\cdots\be(y_{n-1},y_n)\be(y_n,y_1)\quad,\nl
O_n & = & \intl_L\;dy_1\;dy_2\cdots dy_n
 v(y_1)\be(y_1,y_2)\cdots\be(y_{n-1},y_n)v(y_n)\quad,
\eqa
which we call {\em closed strings\/} and {\em open strings,\/}
respectively. We can then concentrate on the one-dimensional case first,
and in more dimensions simply use
\bqa
C_n(s\mbox{-dim}) & = & C_n(1\mbox{-dim})^s\quad,\nl
O_n(s\mbox{-dim}) & = & O_n(1\mbox{-dim})^s\quad.
\eqa
Keeping careful account of the combinatorial factors, we have
\bqa
\xp{N^mD_2^m} & = & \suml_{p_{1,2,\ldots}}\suml_{q_{1,2,\ldots}}
{m!\over p_1!p_2!\cdots q_1!q_2!\cdots}\;{(2r)!\over2^rr!}\nl
& &\hphantom{XXX}\times\;
\prol_{l\ge1}\left({2^{l-1}\over l}C_l\right)^{p_l}
\prol_{n\ge1}\left(-2^{n-1}O_n\right)^{q_n}\quad,
\eqa
with the constraints
\bqa
m & = & p_1+2p_2+3p_3+\cdots+q_1+2q_2+3q_3+\cdots\quad,\nl
r & = & q_1+q_2+q_3+\cdots\quad.
\eqa
The moment-generating function is, therefore, 
\bqa
G_2(z) & \equiv & \suml_{m\ge0}{z^m\over m!}\xp{N^mD_2^m}
 = {\exp(\psi(z))\over\sqrt{\chi(z)}}\quad,\nl
\psi(z) & = & \suml_{n\ge1}{(2z)^n\over2n}C_n\quad,\nl
\chi(z) & = & 1 + \suml_{n\ge1}(2z)^nO_n\quad.
\eqa
So, provided we can compute the closed and open strings for a given
problem class, we can compute the probability density $f_2(x)$ for
$ND_2$ to have the value $x$, by Fourier transform:
\bq
f_2(x) =  {1\over2\pi}\intl_{-\infty}^{\infty}
\;dz\;e^{-izx}G_2(iz)
= {1\over\pi}\intl_0^{\infty}dz\;
\real\left(e^{-izx}G_2(iz)\right)\quad,
\eq
where we have used $G_2(z^{\ast})=G_2(z)^{\ast}$, since $G_2(z)$
is real for $z$ real and sufficiently small.
Finally, by taking various derivatives of $G_2(z)$ at $z=0$,
we can compute the moments of the $ND_2$ distribution, and find
\bqa
\mbox{mean} & = & C_1 - O_1\quad,\nl
\mbox{variance} & = & 2(C_2-2O_2+O_1^2)\quad,\nl
\mbox{skewness} & = & 
\sqrt{8}\;{C_3-3O_3 + 3O_2O_1 - O_1^3\over(C_2-2O_2+O_1^2)^{3/2}}\quad,\nl
\mbox{kurtosis} & = & 
12\;{C_4 - 4O_4 + 4O_3O_1 + 2O_2^2 - 4O_2O_1^2 + O_1^4\over
(C_2-2O_2+O_1^2)^2}\quad.
\label{moments}
\eqa

\subsection{The Wiener case}
We shall now concentrate on the Wiener problem class.
As usual we shall start with $s=1$. In this case, we have
\bq
\be(y_1,y_2) = \min(y_1,y_2)\quad\;,\quad\;v(y) = y\quad.
\eq
The strings can be computed most easily by first defining
\bq
A_n(x,y) = \intu\;dy_2\;dy_3\cdots dy_n
 \be(x,y_2)\be(y_2,y_3)\cdots\be(y_{n-1},y_n)\be(y_n,y)\quad,
\eq
so that
\bq
C_n = \intu\;dx\;A_n(x,x)\;\;,\;\;
O_n = A_{n+1}(1,1)\quad,
\eq
where the second line holds only in this, the Wiener, case.
The objects $A_n$ obey a recursion relation:
\bqa
A_1(x,y) & = & \be(x,y)\quad,\nl
A_n(x,y) & = & \intu\;dz\;A_{n-1}(x,z)\be(z,y)\;\;,\;\; n=2,3,\ldots\quad.
\eqa
They can most easily be computed using a generating function 
that itself satisfies the obvious integral equation:
\bq
F(t;x,y) = \suml_{n\ge1}A_n(x,y)t^n\;\;,\;\; = t\intu\;dz\;F(t;x,z)\be(z,y) + t\be(x,y)\quad,
\eq
from which we shall establish a differential equation. Since
\bq
\min(x,0) = 0\;\;,\;\;
{\partial\over\partial y}\min(x,y) = \theta(x-y)\;\;,\;\;
{\partial^2\over\partial y^2}\min(x,y) = -\delta(x-y)\quad,
\eq
we can conclude that
\bq
{\partial^2\over\partial y^2}F(t;x,y) = - tF(t;x,y) - t\delta(x-y)\quad,
\label{partialdiffeqn}
\eq
with the boundary conditions
\bq
F(t;x,0) = 0\;\;,\;\;
\left[{\partial\over\partial y}F(t;x,y)\right]_{y=0} = 
t\intu\;dz\;F(t;x,z) + t\quad.
\eq
We can solve \eqn{partialdiffeqn} by standard methods, 
and, with $t=u^2$, we have
\bq
F(t;x,y) = -u\theta(y-x)\sin(uy-ux)
+{u\over\cos u}\cos(u-ux)\sin(uy)\quad,
\label{explicitformforf}
\eq
which is actually symmetric in $x$ and $y$, as it should be.
{}From \eqn{explicitformforf} we can  compute $C_n$ and $O_n$ in a
straightforward manner, since
\bq
\suml_{n\ge1}C_nt^n = \intu\;dx\;F(t;x,x)\;\;,\;\;
\suml_{n\ge1}O_nt^n = {1\over t}F(t;1,1)\quad,
\eq
and we arrive at
\bqa
C_n & = & \left({4\over\pi^2}\right)^n\xi(2n)
    \; = \; {2^{2n-1}\left(2^{2n}-1\right)\over(2n)!}|B_{2n}|\quad,\nl
O_{n-1} & = & 2C_n\quad,\nl
\xi(p) & = & \suml_{n\ge1}{1\over(2n-1)^p} = (1-2^{-p})\zeta(p)\quad,
\eqa
where $\zeta(2n)$ is the Riemann zeta function, and
$B_{2n}$ is the Bernoulli number.
The first few instances are
\bq
C_1={1\over2}\quad,\quad C_2={1\over6}\quad,\quad
C_3={1\over15}\quad,\quad C_4={17\over630}\quad,\quad
C_5={31\over2835}\quad.
\eq
Incidentally, we also get $O_0=1$, in accordance
with the fact that the series expansion of $\chi(z)$
must start with 1. The functions $\psi$ and $\chi$ obtain the form
\bqa
\psi(z) & = & -{1\over2}\suml_{n\ge1}\log\left(
1-{8z\over\pi^2(2n-1)^2}\right)\; = \; -{1\over2}\log\cos\sqrt{2z}\quad,\nl
\chi(z) & = & \suml_{n\ge1}{8\over\pi^2(2n-1)^2-8z}
\; = \; {1\over\sqrt{2z}}\tan\sqrt{2z}\quad.
\label{explicitformsisone}
\eqa
 

\begin{figure}
\begin{center}
  \input{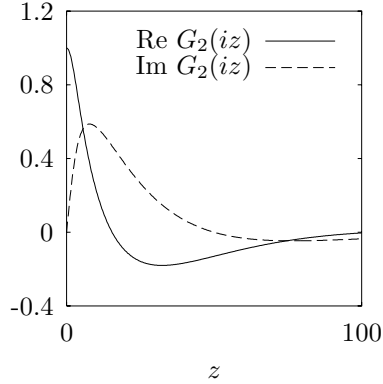}
\end{center}
\caption[.]{Real and imaginary part of $G_2(iz)$ for $s=1$.}
\end{figure}


We can, using the series expansion forms, compute $\psi(iz)$ and 
$\chi(iz)$ to arbitrary accuracy. We might also use the fact that,
in one dimension,
\bq
G_2(z) = \left[{\sqrt{2z}\over\sin\sqrt{2z}}\right]^{1/2}\quad,
\label{niceanalyticform}
\eq
but, what with all the branch cuts this function has in the complex $z$
plane, it is not easy to extend the computation of this more compact
form so that $G_2(z)$ is adequately defined along the whole
imaginary axis; and we shall use the form (\ref{explicitformsisone}).
Note that we may `truncate' the series at large enough $n$, and sum
the remainder in the approximation $z\ll n^2$: this gives, for instance
\bqa
\psi(z) & \sim & {1\over2}z + {1\over6}z^2 
 - {1\over2}\suml_{n=1}^K
\left[\log(1-z_n)+z_n+{1\over2}z_n^2\right]\quad,\nl
\chi(z) & \sim & 1 + {2\over3}z
 + {1\over z}\suml_{n=1}^K{z_n^3\over1-z_n}\quad,\nl
z_n & \equiv & \left({4\over\pi^2}\right){(2z)\over(2n-1)^2}\quad,
\eqa
where $K$ is chosen large enough. In fig.11 we present the
generating function $G_2(iz)$ for positive $z$.
 

\begin{figure}
\begin{center}
  \input{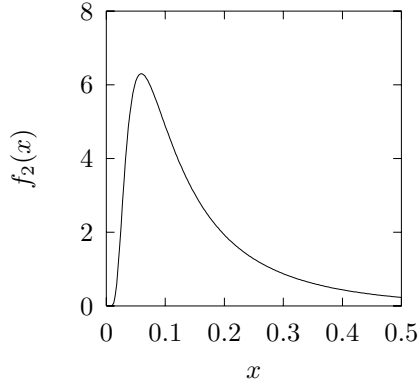}
\end{center}
\caption[.]{Probability density $f_2(ND_2)$, for $s=1$.}
\end{figure}


The Fourier transform now gives us $f_2(ND_2)$ for $s=1$.
Its shape is given in fig.12, for $D_2>0$. We have checked
that $f_2(x)=0$ for negative $x$. The shape of the density is seen to
be skewed, and not to approximate a Gaussian in the large-$N$ limit (in
fact, there is no Central Limit Theorem telling us that it should):
the maximum is somewhat lower than the mean. It can be checked numerically
that $f_2(x)$ is normalized to 1, and the first two moments are
1/6 and 1/20, respectively, in agreement with \eqn{moments}.

\begin{table}[ht]\begin{center}\begin{tabular}{|c|c|c|c|} \hline \hline
$s$ & mean & st.dev. & skewness \\ \hline
   1 &    0.167E+00 &    0.149E+00 &    0.256E+01 \\
   2 &    0.139E+00 &    0.956E-01 &    0.239E+01 \\
   3 &    0.880E-01 &    0.502E-01 &    0.236E+01 \\
   4 &    0.502E-01 &    0.242E-01 &    0.234E+01 \\
   6 &    0.143E-01 &    0.491E-02 &    0.231E+01 \\
   8 &    0.375E-02 &    0.915E-03 &    0.226E+01 \\
  10 &    0.960E-03 &    0.163E-03 &    0.220E+01 \\
  12 &    0.242E-03 &    0.283E-04 &    0.214E+01 \\
  16 &    0.152E-04 &    0.819E-06 &    0.200E+01 \\
  20 &    0.953E-06 &    0.231E-07 &    0.186E+01 \\
  24 &    0.596E-07 &    0.647E-09 &    0.172E+01 \\
  32 &    0.233E-09 &    0.501E-12 &    0.147E+01 \\
  40 &    0.909E-12 &    0.387E-15 &    0.125E+01 \\
\hline \hline \end{tabular}
\caption[.]{Parameters of $f_2(x)$ as a function of dimension.}
\end{center}\end{table}
We now turn to the case of $s$ dimensions. We have to replace
$C_n\to C_n^s$, $O_n\to O_n^s$, and the nice analytic form 
(\ref{niceanalyticform}) of the moment-generating function is lost.
The series form for $\psi$ and $\chi$ can be written in a form
similar to that of \eqn{explicitformsisone}, if we introduce
$P_s(n)$,  the {\em factor multiplicity\/} in $s$ dimensions. This number
counts in how many ways an odd integer $n$ can be written as a product
of $s$ odd integers, including factors 1. In appendix D we show how
this can easily be computed. The functions $\psi$ and $\chi$ then
take the form
\bqa
\psi(z) & \sim & \left({1\over2}\right)^sz + 
\left({1\over6}\right)^sz^2 
 - {1\over2}\suml_{n=1}^KP_s(2n-1)
\left[\log(1-z_n)+z_n+{1\over2}z_n^2\right]\quad,\nl
\chi(z) & \sim & 1 + 2z\left({1\over3}\right)^s
 + {2^s\over 2z}\suml_{n=1}^KP_s(2n-1){z_n^3\over1-z_n}\quad,\nl
z_n & \equiv & \left({4\over\pi^2}\right)^s{(2z)\over(2n-1)^2}\quad,
\eqa
We can now compute the $f_2(x)$ in these more-dimensional
cases. 
In doing so, we must realize that $G_2(z)$ becomes increasingly
flat, and hence $f_2(x)$ increasingly more peaked, with
increasing $s$, as is evident from table 1, where we give
the mean, standard deviation $\sigma$, and skewness of $f_2(x)$
as a function of $s$.
Note that the deviation from a Gaussian, as indicated by
the skewness, decreases with increasing $s$, but does so very
slowly: the skewness decreases by a factor of 10 for an increase in
$s$ by about 113, and attains the value 0.001 only for $s=389$.
For `medium' dimensionality, such as is common in high-energy
phenomenology, we cannot trust the Gaussian approximation, and
the full numerical Fourier transform has to be used.
The results are best presented in the form of the standardized
variable
\bq
\xi = {x-\xp{x}\over\sigma}\quad.
\label{def:xi}
\eq                                       
In fig.13 we plot $\sigma f_2(x)$ as a function of $\xi$ for
various dimensions. For comparison, we also give the shape of
the standard normal distribution under the same plotting convention.


\begin{figure}
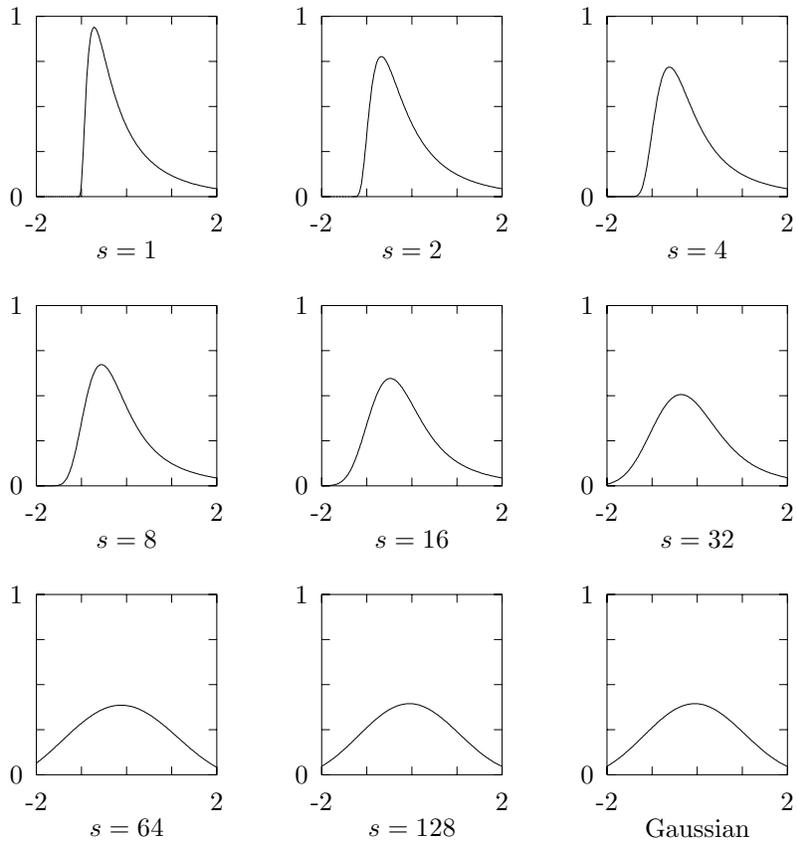

\begin{center}
\mbox{
  \input{figf2_s=1_new}\hspace{-7.5mm}
  \input{figf2_s=2_new}\hspace{-7.5mm}
  \input{figf2_s=4_new}}
\mbox{
  \input{figf2_s=8_new}\hspace{-7.5mm}
  \input{figf2_s=16_new}\hspace{-7.5mm}
  \input{figf2_s=32_new}}
\mbox{
  \input{figf2_s=64_new}\hspace{-7.5mm}
  \input{figf2_s=128_new}\hspace{-7.5mm}
  \input{figf2_gauss_new}}
\end{center}
\caption[.]{Probability density $\sigma f_2(\xp{x}+\sigma\xi)$
 for the Wiener problem set, as a function of the
 standardized variable $\xi$, for dimensions from 1 to 128.
 The last plot is the normal distribution with the same convention.}
\end{figure}


For comparison purposes, the confidence levels implied by $f_2(x)$
for a given dimensionality are relevant, as we have discussed.
We therefore give, in table 2, a number of quantiles for a number
of different dimensions. 
\begin{table}[ht]\begin{center}
\begin{tabular}{|l|c|c|c|c|c|c|c|c|} 
\hline\hline
quantile & \multicolumn{8}{|c|}{dimensionality} \\
       & 1    & 2    & 4    & 8    & 16   & 32   & 64   & $\infty$ \\ 
\hline
0.1 \% & -1.00& -1.15& -1.27& -1.39& -1.66& -2.06& -2.53& -3.09\\
1 \%   & -0.95& -1.06& -1.14& -1.23& -1.40& -1.67& -1.97& -2.33\\
5 \%   & -0.87& -0.94& -0.98& -1.03& -1.12& -1.28& -1.46& -1.64\\
10 \%  & -0.81& -0.86& -0.88& -0.90& -0.96& -1.06& -1.17& -1.28\\
50 \%  & -0.32& -0.29& -0.28& -0.26& -0.22& -0.16& -0.09&  0.00\\
90 \%  & 1.21 & 1.22 & 1.21 & 1.21 & 1.21 &  1.22&  1.25&  1.28\\
95 \%  & 1.98 & 1.96 & 1.95 & 1.94 & 1.90 &  1.83&  1.74&  1.64\\
99 \%  & 3.87 & 3.80 & 3.78 & 3.74 & 3.63 &  3.36&  2.93&  2.33\\
99.9 \%& 6.72 & 6.56 & 6.54 & 6.43 & 6.25 &  5.70&  4.75&  3.09\\
\hline
        &7.6E-5&6.3E-5&6.0E-5&5.1E-5&4.1E-5&1.7E-5&  2E-6&    0 \\
\hline \hline \end{tabular}
\caption[.]{Quantiles of the standardized distribution of the
quadratic discrepancy for the Wiener problem set, for various dimensions. 
The infinite dimensionality corresponds to the normal distribution.
The last line gives the accuracy of the total integral.}
\end{center}
\end{table}
Note that the 50\% quantile corresponds
to the mode of the distribution, which (due to the skewness) is
not necessarily equal to the mean value, which for the standardized
variable $\xi$ is of course zero. To have an impression of the
numerical accuracy of the quantile determination, we also give the
difference between 1 and the actual numerical value of the total integral,
which ideally should be zero. The behaviour of the quantiles with
dimension shows that the case $s=1$ is not really typical, since
the distribution for $s=2$ appears to be much wider. This can be
understood somewhat if we realize that $f_2(x)$ vanishes for
negative $x$: if $\sigma$ is comparable to $\xp{x}$, this 
forces the low quantiles to be relatively close to zero.

\subsection{Orthonormal case}
We now turn to the case of a problem set defined in terms of
orthonormal functions, such as the Fourier or Walsh set.
In these cases, there is no problem with factorization in higher
dimensions. For $s=1$, we have
\bq
\al(y_1,y_2) = \suml_{n\ge1}\si_n^2u_n(y_1)u_n(y_2)\quad;
\eq
from which it is trivial to prove that
\bq
Z_m = \suml_{n\ge1}\si_n^{2m}\quad.
\eq
Now, we may use $C_n=Z_n$, and $O_n=0$, to construct the
moment-generating function:
\bqa
G_2(z) & = & \exp(\psi(z))\quad,\nl
\psi(z) & = & -{1\over2}\suml_{n\ge1}\log\left(1-2z\sigma_n^2\right)\nl
 & = & -{1\over2}\suml_{n\ge1}^K\left[\log\left(1-2z\sigma_n^2\right)
 + 2z\sigma_n^2 + 2z^2\sigma_n^4\right]
 + z\suml_{m\ge1}\sigma_n^2 + z^2\suml_{m\ge1}\sigma_n^4\quad,\nl
\eqa
where we have also indicated the proper way to evaluate it.
In the more-dimensional case, we might assume that
\bq
\sigma_{\vec{n}} = \sigma_{(n_1,n_2,\ldots,n_s)} = 
\sigma_{n_1}\sigma_{n_2}\cdots\sigma_{n_s}\quad,
\eq
and obtain an analogous formula. An attractive choice would be
to have
\bq
\sigma_{n} = {1\over2n-1}\quad,
\eq
so that we may use the factor multiplicity $P_s(n)$ again:
\bqa
G_2(z) & = & \exp(\psi(z))\quad,\nl
\psi(z) & = & -{1\over2}\suml_{n\ge1}
 P_s(n)\log\left(1-2z\sigma_n^2\right)\nl
 & = & -{1\over2}\suml_{n\ge1}^K
 P_s(n)\left[\log\left(1-2z\sigma_n^2\right)
 + 2z\sigma_n^2 + 2z^2\sigma_n^4\right]
 + z\xi(2)^s + z^2\xi(4)^s\quad,\nl
\eqa

\section{Discrepancy Calculations}

\subsection{Extreme Discrepancy}
Numerical calculation of the extreme discrepancy of a particular
point set requires first calculating the local discrepancy at a 
position in the hypercube, then finding the maximum of that local 
discrepancy over the hypercube.  
The local discrepancy is calculated in time $N s$, 
where $N$ is the number of points and $s$ the dimensionality,
since it involves essentially counting the number of points all of whose
coordinates are less than the position being considered. 
For high dimensionality some time can be saved because as soon 
as one coordinate of a point is greater than the coordinate of the 
position, it is not necessary to test the others, so the actual time
would behave more like $N \sqrt{s}$.  
The major problem is of course finding the maximum of this function.
The local discrepancy is discontinuous not only at every point,
but also at every position of which any coordinate is the coordinate
of a point.  This means there are $N^s$ candidate positions where the
maximum could potentially occur.  
Again a clever program can save some time because some of these
positions are not in fact possible maxima, but still the overall 
computational complexity can be expected to behave approximately as
$N \sqrt{s} N^s$, which makes it prohibitive to calculate for 
large point sets in high dimensions. Practical limits are around ten
points in ten dimensions or 100 points in five dimensions, 
which are too small to be of interest for real calculations.

\subsection{Exact Quadratic Discrepancy}
Although it may appear that quadratic discrepancy is more complicated
than extreme discrepancy because it is defined as an integral over the
hypercube, in fact the integrals involved are only step functions
and it is possible to perform considerable simplification to reduce 
the calculation time. 
Perhaps the best way to explain how the calculation works is simply
to give the Fortran code:
 
\begin{verbatim}
      subroutine discd2(ntot,ns,x,d2)
**********************************************************
* compute the quadratic discrepancy for the point set
* x(i,mu), i=1,2,...,ntot    mu=1,2,...,ns
* the output is the array d2(n), n=1,2,...,ntot
* and gives the discrepancy * n^2 for the first n points
**********************************************************          
      implicit real*8(a-h,o-z)
      dimension x(ntot,ns),d2(ntot)
* initialize a few constants
      c2 = (1d0/2d0)**ns
      c3 = (1d0/3d0)**ns
      bn = 0
* start loop over number of points
      do 999 n = 1,ntot
        if(mod(n,100).eq.0) print *,'doing',n,'...'
* compute b(n) and a(n,n) for the new point
        a = 1.d0
        b = 1.d0
        do 1 mu = 1,ns
          a = a*(1.d0-x(n,mu))
          b = b*(1.d0-x(n,mu)**2)
    1   continue
        b = c2*b
* update running value of sum_b
        bn = bn+b
* case n=1
        if(n.eq.1) then
          d2(n) = a - 2*bn + c3
* case n>1
        else
* sum the a(i,n) for i=1 to n-1
          an = 0.d0
          do 3 i = 1,n-1
            temp = 1.d0
            do 2 mu = 1,ns
              temp = temp*(1.d0-dmax1(x(i,mu),x(n,mu)))
    2       continue
            an = an+temp
    3     continue
* give d2(n) for n>1 by relating it to d2(n-1)
          d2(n) = d2(n-1) + 2*an + a - 2*bn -2*(n-1)*b + (2*n-1)*c3
        endif
* end of loop over n
  999 continue
       end
\end{verbatim}

Note that we used double precision to calculate the discrepancy,
even though we generated the points in single precision.
This is necessary since considerable precision is lost in the big sums,
and we found that in single precision only about one or two digits of
accuracy remained for the larger point sets.

\subsection{Quadratic discrepancy by Monte Carlo} 
Since the quadratic discrepancy is a multidimensional integral, it
can be estimated using a Monte Carlo approximation.  
If the desired accuracy is, for example 5\% , 
we found Monte Carlo to be faster than 
the exact calculation for more than about 50,000 points. 
We used ordinary pseudorandom numbers for these calculations 
for fear that correlations between the two quasi-random point sets 
would produce unreliable results. We repeated some Monte Carlo 
calculations using the exact algorithm to verify that the agreement
was good.   
 
\subsection{Numerical Results}
 
We have calculated the quadratic discrepancy for four different 
quasi-random generators:
Richtmyer, van der Corput-Halton, Sobol' and Niederreiter-base-two, 
all from one to twenty dimensions.  
Exact calculations were made for 
1-100,000 points, and Monte Carlo calculations for 
100,000 and 150,000 point sets.  

The programs used to generate the Richtmyer points and van der Corput 
points were written by us, and work in any number of dimensions 
limited only by some dimension statements and the table of prime numbers.
The Sobol' and Niederreiter programs were kindly provided by
Paul Bratley and Bennet Fox.  Their Sobol' program is implemented 
for up to 40 dimensions, and we have extended their 
Niederreiter-base-two programs to work up to 25 dimensions. 

We have consumed weeks of computer time producing dozens of plots of 
quadratic discrepancy as a function of $N$  for the four quasi-random 
generators and also some good pseudorandom generators, 
in all dimensions from one to 20.
It would be fastidious to show all these plots here, especially since 
many of them are rather similar and most of the behaviour can be 
summarized in some more revealing plots as we later discovered.
However, we do show here a few typical plots as well as those which 
we found the most surprising.


\begin{figure}
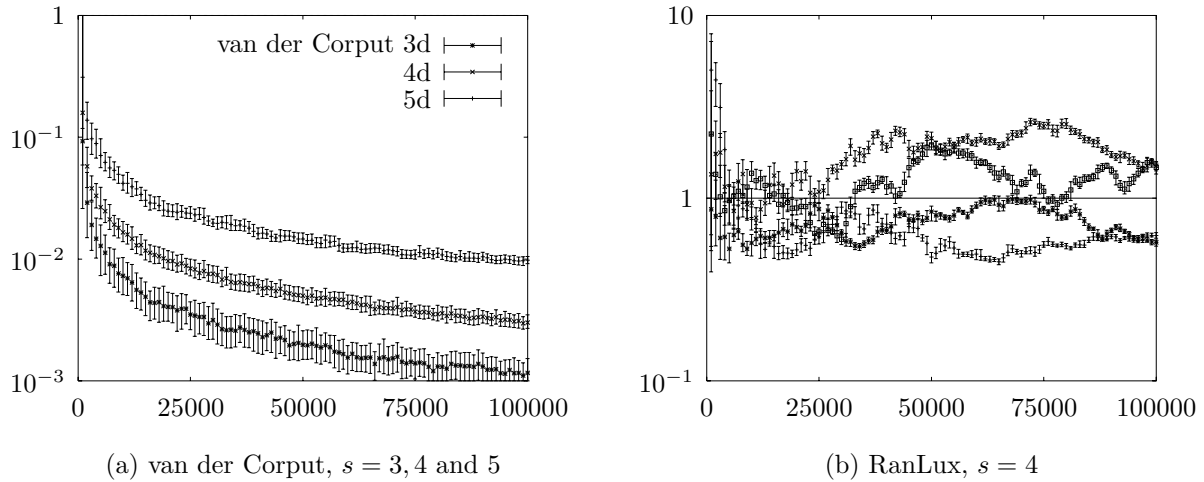

\begin{center}
\mbox{\hspace{-3.5cm}
  \input{figD2C345.tex} \hspace{-1cm}
  \input{figD2L4.tex}}
\end{center}
\vspace{0.75cm}
\caption[.]{The quadratic discrepancy for up to 100~000 points for
(a) the van der Corput generator in 3, 4 and 5 dimensions; and 
(b) four samples of random points in 4 dimensions.}
\label{D2C345L4}
\end{figure}


Figure~\ref{D2C345L4} shows some examples of quadratic discrepancy
calculated for random and quasi-random point-sets up to 100~000 points.
In these plots, the error bars give the total range of quadratic 
discrepancies calculated for all the 1000 values of $N$ in the
range around the plotted value. Thus the tops and bottoms of the 
error bars give the envelope of the full plot, had we plotted a
point for each value of $N$.  
Subfigure (a) shows a typical quasi-random behaviour for low dimensionality, 
in this case the van der Corput generator for 3, 4 and 5 dimensions.
One sees that apart from very small point-sets, there is a considerable
improvement over the expectation for random points (defined here as =1.0),
and this improvement increases (faster convergence) with increasing $N$,
as expected from the theoretical results. 
However, with increasing dimension, the improvement over random points
decreases.
For comparison, figure~\ref{D2C345L4}(b) shows the same 
quadratic discrepancy for four typical random point sets.
Note the different scale, however, as these discrepancies are much larger
than those of quasi-random points. 


\begin{figure}
\begin{center}\hspace{-3.5cm}
  \input{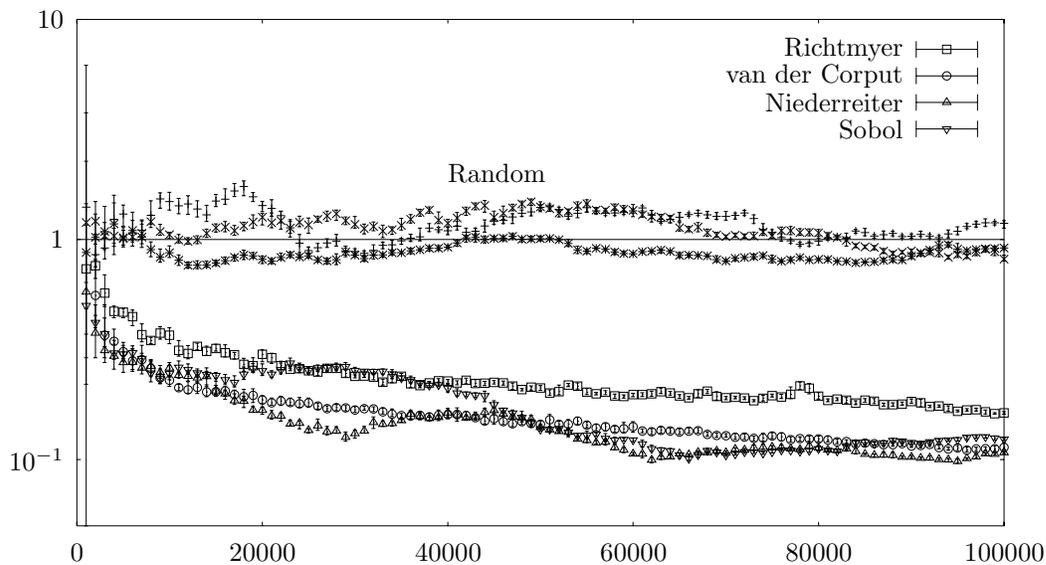}
\end{center}
\caption[.]{The quadratic discrepancy for up to 100~000 points for
both random and quasi-random points in 8 dimensions. 
Three different random point samples are compared with our four
quasi-random generators which all have much smaller discrepancy.}
\label{D2all8}
\end{figure}


Figure~\ref{D2all8} shows the discrepancy in eight dimensions
for both random and quasi-random points.
Our normalized values of quadratic discrepancy for 
three random sequences
are of course close to one as they must be.
All the four quasi-random sequences show a much lower discrepancy, 
between a factor five and ten lower for 50 to 100~000 points. 
For more than 40~000 points, the Richtmyer generator is not quite 
as good as the other three quasi-random generators. Note that
these results are obtained for dimension 8. The results for other
low dimensionalities are qualitatively similar.


\begin{figure}
\begin{center}\hspace{-1cm}
  \input{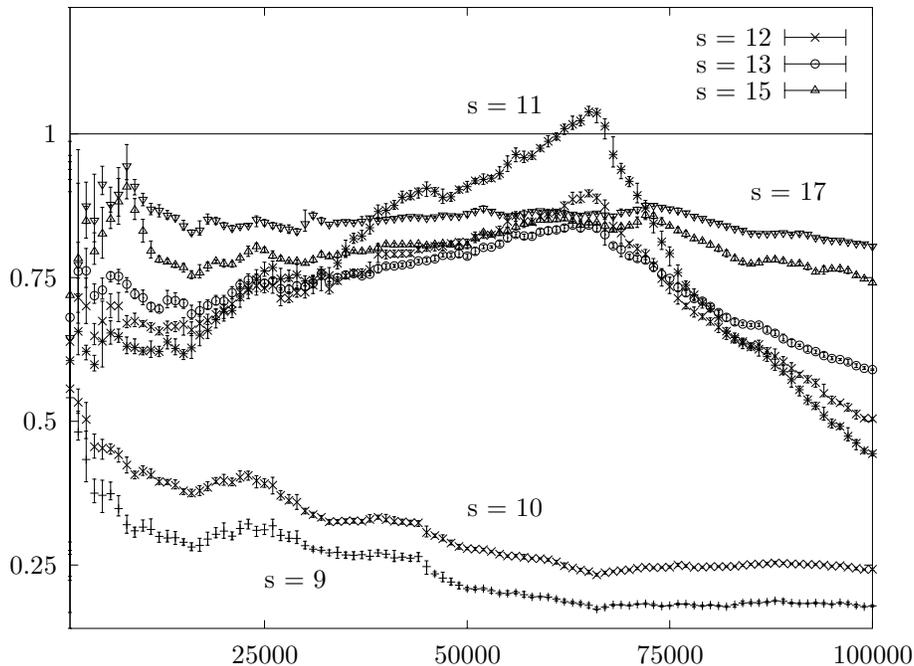}
\end{center}
\caption[.]{The quadratic discrepancy for up to 100~000 points for
the Sobol' sequence in dimensions 9 to 17.}
\label{weirdsobol}
\end{figure}


It must be realized, however, that for higher dimensionalities
surprises may be in store. As an illustration, we give in 
figure~\ref{weirdsobol} the discrepancy for the Sobol' sequence in 
dimensions 9 to 17, as a function of the number of points. 
It is seen that around 70,000 points the discrepancy it not small at all, 
and is actually worse than
random for $s=11$; after this bump, however, the discrepancy quickly
returns to a very small value. Although such a behaviour can
presumably be well understood from the underlying algorithm, it must
be taken as a warning that for quasi-random sequences, an increase
in the number of points does {\em not\/} automatically lead
to an improvement. Note that this behaviour is not necessarily
the same as the well-known `irregularity of distribution', since
in figure~\ref{weirdsobol} it is actually the envelope of the 
discrepancy curve that shows a global, long-term rise (for $s=11$, 
running from $N\sim30,000$ to $N\sim70,000$, a much larger scale 
than ought to be expected for {\em local\/} increases in discrepancy).


\begin{figure}
\begin{center}
\mbox{\hspace{-2.5cm}
  \input{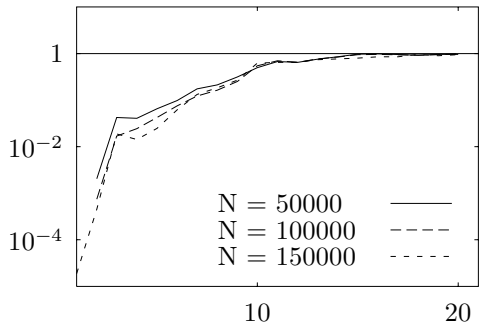}\hspace{-1.5cm}
  \input{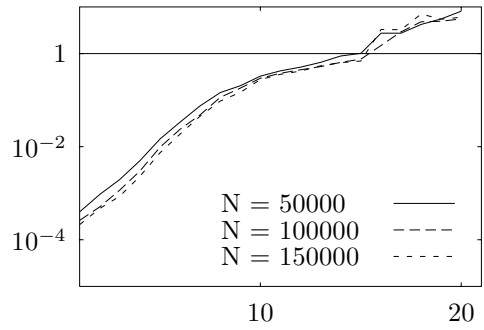}}
\vspace{-0.3cm}
\end{center}
\begin{center}
\mbox{\hspace{-2cm}
  \input{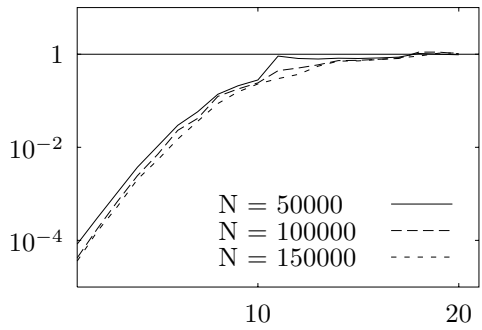}\hspace{-1.5cm}
  \input{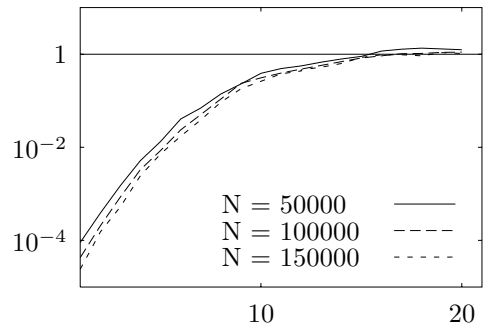}}
\vspace{0.3cm}
\end{center}
\caption[.]{The quadratic discrepancy, normalized to the quadratic 
discrepancy for truly random numbers for N=50~000, 100~000 and 150~000
points as a function of dimensionality for four different quasi-random
generators.}
\label{D2all}
\end{figure}


Figure~\ref{D2all} shows a summary of the 
quadratic discrepancies for all our quasi-random generators.
As usual, all discrepancy values are normalized to 
the expectation for a truly random point set. 
Since we found the behaviour with respect to $N$ rather uninteresting 
for most cases, we have instead plotted here the discrepancy 
for a few fixed values of $N$ as a function of the dimensionality $s$.
This is considerably more revealing, and in fact the first striking 
aspect is the resemblance between the behaviours of all the different
generators.

Now we see that the dominant effect is that all generators are good 
for small $s$ and all discrepancies rise sharply as a function of $s$,
finally approaching the random expectation around $s=15$ for $N=150~000$.  

The van der Corput generator becomes much worse than random
above $s=15$, whereas the other discrepancies remain rather 
close to random up to $s=20$.  Surprisingly, the only one that stays 
better than random up to $s=20$ is the Richtmyer.


\begin{figure}
\begin{center}
\mbox{\hspace{-1.8cm}
  \input{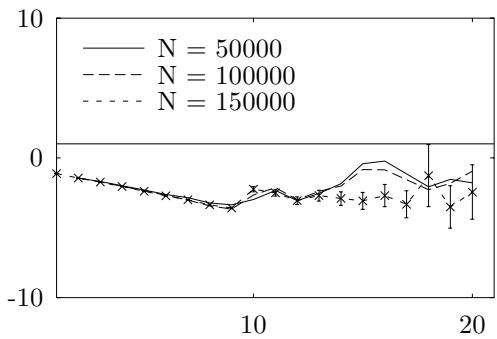}\hspace{-1cm}
  \input{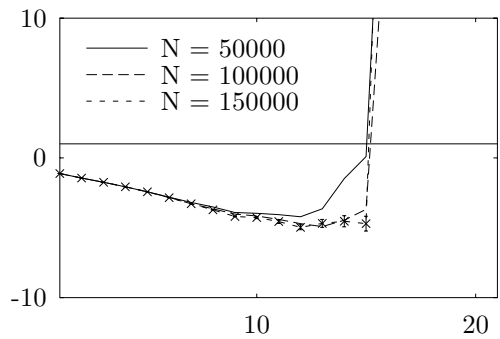}}
\vspace{0.2cm}
\end{center}
\begin{center}
\mbox{\hspace{-1.3cm}
  \input{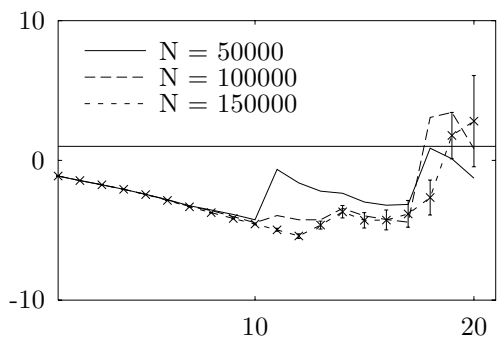}\hspace{-1cm}
  \input{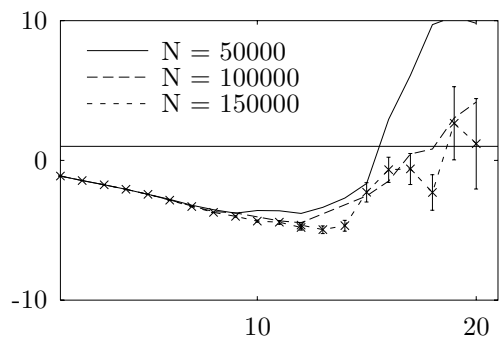}}
\vspace{0.3cm}
\end{center}
\caption[.]{The standardized quadratic discrepancy $\xi$ for N=50~000, 100~000 a
nd 150~000 
points as a function of dimensionality for four different quasi-random
generators. For $N=150~000$, the Monte Carlo error-estimate is given.}
\label{standD2all}
\end{figure}


In Figure~\ref{standD2all} we plot the discrepancies in terms of the 
variable $\xi$ of \eqn{def:xi}. This representation
ought to inform us about how `special' the quasi-random sequences are
compared to truly random ones, since on this scale 95\%\  of all truly 
random sequences will fall between -2 and +2.
Now we see that the lower discrepancy of the Richtmyer sequence
compared to random sequences is really significant up to 20 dimensions,
provided one uses more than 100~000 points.
The poor behaviour of the Van~der~Corput sequences above $s=15$
is highly significant, but the other generators look much like random 
ones between $s=18$ and $s=20$.
This provides additional motivation to study the Richtmyer generator,
which is easily implementable,
and in particular to look for optimal constants $S_\mu$. 

It should be noted, that the discrepancy improves in general 
with increasing numbers of points.
It might be conjectured, that asymptotic behaviour of sequences' 
discrepancy will, for $s>10$, only become evident for a larger number
of points. 

\section{Conclusions}

We have reviewed a number of arguments that suggest that the use of
quasi-random numbers in multi-dimensional integration may lead to an
improved error over classical Monte Carlo.
The central notion in this context is that of discrepancy.
We have shown, that there exist a relation between the definition of
a discrepancy and the class of integrands of which our particular 
integrand is supposed to be a typical member.
We have discussed several such classes and derived the appropriate induced
discrepancies.

Another important aspect of discrepancy is the fact, that for a 
quasi-random point set it ought to be smaller than for a truly random
point set. We have therefore studied the distribution of the value 
of discrepancy for truly random points.

Finally,  we have computed the quadratic Wiener discrepancy for a 
number of quasi-random point sets. 
For all the quasi-random generators tested, the discrepancy, normalized
to that of truly random numbers increases with increasing dimensionality 
and approaches that of truly random point sets around $s=15$.
If the quadratic discrepancy is plotted in terms of the standardized 
variable $\xi$, it is seen that the quadratic discrepancy for the 
quasi-random point sets is indeed significantly better than that of 
a typical member of the ensemble
of random point sets for dimensions up to $s\approx12$ for $N=100~000$ 
to $s\approx15$ for $N=150~000$. 
Only one quasi-random generator (the Richtmyer) remains significantly
better than random up to the highest dimension we calculated (twenty).
For higher dimensions, it is conjectured that
the asymptotic regime for the discrepancy is approached only for a much
higher number of points $N$, which we did not cover in our computations.

\newpage
\section*{Appendix A: Van der Corput discrepancy}
Here we present some considerations on the behaviour of the
van der Corput sequence in base 2, that can be obtained in a straightforward
manner. We consider the extreme discrepancy for the finite $N$-member
point set $x_k=\phi_2(k)$, with $k=0,1,2,\ldots,N-1$, and we define
\bq
d(N) = ND_{\infty}(N)\quad.
\eq
We trivially have that $d(N)\ge1$; more precisely, by inspection
of the behaviour (given in fig.1) we observe the following recursion:
\bqa
d(0) & = & 0\quad,\nl
d(N) & = & \left\{\begin{tabular}{ll}
 $d({N\over2})$ & \mbox{if $N$ is even,} \\
 $\vphantom{X}$ & \\
 ${1\over2}\left(1+d({N-1\over2})+d({N+1\over2})\right)$
  & \mbox{if $N$ is odd.}
 \end{tabular}\right.
\label{corputdisrecur}
\eqa
This recursion sheds some light on how the discrepancy depends on
the binary properties of $N$, and
in this manner we can compute $d(N)$ very quickly even for extremely
large $N$. Now, we find the maximal discrepancy in the following way:
in each interval between $N=2^{p-1}$ and $N=2^p$, we search for
the maximum of $d(N)$. This is reached for two values of $N$, and
we concentrate on the smallest one of these, which we denote by
$\nu_p$: the corresponding value $d(N)$ we denote by $d_p$.
Again by inspection, we find the following recursions:
\bq
\nu_p = 2\nu_{p-1} + (-1)^p\;\;,\;\;
d_p   = {1\over2}\left(d_{p-1}+d_{p-2}+1\right)\quad,
\eq
with starting values $\nu_1=1$, $d_1=1$, and $d_2=3/2$. These
relations are easily solved to give closed formulae for $\nu_p$ and $d_p$:
\bq
\nu_p = {1\over3}\left(2^{N+1}+(-1)^p\right)\;\;,\;\;
d_p   = {1\over9}\left(3N+7-(-2)^{1-N}\right)\quad.
\eq
Now, as $N$ becomes large, we can easily eliminate $p$, and we
find the asymptotic behaviour
\bq
d_p \sim {\log\nu_p\over\log8}\quad,
\eq
which agrees nicely with \eqn{corputdisconedim}. 
For other bases, recursion relations similar to \eqn{corputdisrecur} 
ought to exist, but we have not pursued this.

\section*{Appendix B: completeness of the Fourier set}
We present a short and simple proof of the completeness of the
orthonormal functions $u_n(x)$ introduced in \eqn{fourierbase}.
To this end, consider the sum
\bqa
\Delta_N(x,y) & = & \suml_{n=0}^{2N}u_n(x)u_n(y)\nl
& = & 1 + 2\suml_{n=1}^N\left[\cos2\pi nx \cos2\pi ny +
\sin2\pi nx \sin2\pi ny\right]\nl
& = & 1 + 2\suml_{n=1}^N\cos2\pi n(x-y)\nl
& \equiv & \Delta_N(x-y)\quad.
\label{fouriernsum}
\eqa
We consider the limit
\bq
\Delta(\xi) = \lim_{N\to\infty}\Delta_N(\xi)\quad.
\eq
Obviously, this diverges for $\xi=0$, and it remains to prove that
we get zero if $\xi$ is not zero. But we have, for such $\xi$,
\bqa
\Delta(\xi) & = & 1 + \suml_{n>0}\real\exp(2i\pi\xi n)\nl
& = & 1 + 2\real\suml_{n>0}(e^{2i\pi\xi})^n\nl
& = & 1 + 2\real{e^{2i\pi\xi}\over1-e^{2i\pi\xi}}\quad,
\eqa
which is indeed zero. Therefore, we have
\bq
\Delta(\xi) = \delta(\xi)\quad,
\eq
the Dirac delta distribution. Note that we have to assume that
the sum over $n$ in \eqn{fouriernsum} has to have an upper limit
$2N$ rather than $N$: the way in which the limit is approached is
important.
Using the property that for a continuous function $f(x)$,
\bqa
f(x) & = &  \int\;dy\;\delta(x-y)\;f(y)\nl
& = & \suml_{n\ge0}u_n(x)\int\;dy\;u_n(y)f(y)\quad,
\eqa
we immediately get the decomposition of $f(x)$ into the orthonormal
base.

\section*{Appendix C: completeness of the Walsh set}
We also prove that the set of all Walsh functions is complete.
Let us first note that
\bq
W_n(x) W_n(y) = W_n(\xi)\quad,
\eq
where $\xi$'s binary expansion $0.\xi_1\xi_2\xi_3\cdots$
 is the bitwise {\tt XOR} of the
binary expansions of $x$ and $y$, that is, $\xi$ has zeroes
in all positions except those where the binary digits of $x$ and $y$
are different. We now define
\bqa
\Delta_N(x,y) & = & \suml_{n=0}^{2^{N+1}-1} W_n(x) W_n(y)\nl
& = & \suml_{n=0}^{2^{N+1}-1} W_n(\xi)\nl
& = & \suml_{n_1=0}^1(-1)^{\xi_1n_1}\suml_{n_2=0}^1(-1)^{\xi_2n_2}\cdots
\suml_{n_{N}=0}^1(-1)^{\xi_Nn_N}\quad.
\label{walshnsum}
\eqa
Now, if $n$ runs from 0 to $2^{N+1}-1$, all its binary digits 
$n_{1,2,\ldots,N}$ will be 0 and 1 an equal number of times.
Therefore, each factor in the sum (\ref{walshnsum}) will evaluate
to zero unless its corresponding digit of $\xi$ is also zero.
Hence, $\Delta_N(x,y)$ is zero unless the binary digits of $x$
and $y$ agree up to and including the $N^{\mbox{{\small th}}}$ 
place, and in that
case it will be $2^N$. Taking the limit $N\to\infty$ we again
recover the Dirac delta distribution. As in the Fourier case, the way
we take the limit turns out to be relevant.

\section*{Appendix D: The function $P_s(n)$}
\begin{table}[ht]
\begin{center}
\begin{tabular}{|r|r|r|r|r|r|r|r|r|r|r|}
\hline \hline
n & \multicolumn{10}{|c|}{$s$} \\
    &    1 &    2 &    3 &    4 &    5 &    6 &    7 &    8 &    9 &   10 \\ 
\hline
  1 &    1 &    1 &    1 &    1 &    1 &    1 &    1 &    1 &    1 &    1 \\
  3 &    1 &    2 &    3 &    4 &    5 &    6 &    7 &    8 &    9 &   10 \\
  5 &    1 &    2 &    3 &    4 &    5 &    6 &    7 &    8 &    9 &   10 \\
  7 &    1 &    2 &    3 &    4 &    5 &    6 &    7 &    8 &    9 &   10 \\
  9 &    1 &    3 &    6 &   10 &   15 &   21 &   28 &   36 &   45 &   55 \\
 11 &    1 &    2 &    3 &    4 &    5 &    6 &    7 &    8 &    9 &   10 \\
 13 &    1 &    2 &    3 &    4 &    5 &    6 &    7 &    8 &    9 &   10 \\
 15 &    1 &    4 &    9 &   16 &   25 &   36 &   49 &   64 &   81 &  100 \\
 17 &    1 &    2 &    3 &    4 &    5 &    6 &    7 &    8 &    9 &   10 \\
 19 &    1 &    2 &    3 &    4 &    5 &    6 &    7 &    8 &    9 &   10 \\
 21 &    1 &    4 &    9 &   16 &   25 &   36 &   49 &   64 &   81 &  100 \\
 23 &    1 &    2 &    3 &    4 &    5 &    6 &    7 &    8 &    9 &   10 \\
 25 &    1 &    3 &    6 &   10 &   15 &   21 &   28 &   36 &   45 &   55 \\
 27 &    1 &    4 &   10 &   20 &   35 &   56 &   84 &  120 &  165 &  220 \\
 29 &    1 &    2 &    3 &    4 &    5 &    6 &    7 &    8 &    9 &   10 \\
 31 &    1 &    2 &    3 &    4 &    5 &    6 &    7 &    8 &    9 &   10 \\
 33 &    1 &    4 &    9 &   16 &   25 &   36 &   49 &   64 &   81 &  100 \\
 35 &    1 &    4 &    9 &   16 &   25 &   36 &   49 &   64 &   81 &  100 \\
 37 &    1 &    2 &    3 &    4 &    5 &    6 &    7 &    8 &    9 &   10 \\
 39 &    1 &    4 &    9 &   16 &   25 &   36 &   49 &   64 &   81 &  100 \\
 41 &    1 &    2 &    3 &    4 &    5 &    6 &    7 &    8 &    9 &   10 \\
 43 &    1 &    2 &    3 &    4 &    5 &    6 &    7 &    8 &    9 &   10 \\
 45 &    1 &    6 &   18 &   40 &   75 &  126 &  196 &  288 &  405 &  550 \\
\hline \hline
\end{tabular}
\caption[.]{Values of $P_s(n)$ for low values of $s$ and $n$.}
\end{center}
\end{table}
In this appendix we discuss the function $P_s(n)$, which counts
in how many ways an odd integer $n$ can be written as a product of
$s$ integers (including ones):
\bq
P_s(n) = \suml_{n_{1,2,\ldots,s}\ge1}\delta_{n_1n_2\cdots n_s\;,\;n}\quad.
\eq
Obviously, $P_1(n)=1$, and we have the recursion
\bq
P_{s+1}(n) = \suml_{m\ge1}P_s(m)\delta_{n\mod m\;,\;0}\quad.
\label{factorrecursion}
\eq
Note that the fact that $n$ is odd automatically restricts these sums
to the odd integers only. The recursion (\ref{factorrecursion})
can trivially be implemented to give us $P_s(n)$ for very large values
of $s$ and $n$. In the table we give its first few values.
Note the irregular behaviour of $P_s(n)$ with $n$: for $n$ prime it
equals $s$, while new maxima are encountered whenever $n$ is a product
of the first consecutive odd numbers, or a triple of that number.
Then, it can become quite large: for instance, $P_{10}(945) = 22000$.
We need, however, not worry about the convergence of series like
those giving $\psi(z)$ and $\chi(z)$; for, we have
\bq
\suml_{n\ge1}P_s(2n-1){1\over(2n-1)^x} =
\left[\suml_{n\ge1}{1\over(2n-1)^x}\right]^s = \xi(x)^s\quad,
\eq
and sums like these are finite whenever $x>1$.
The function $P_s(n)$ {\em without\/} the restriction to odd integers
is discussed, for instance, by Hardy and Wright \cite{hardy}.

\section*{Acknowledgments}
We are grateful to Paul Bratley and Bennet Fox for supplying several 
programs for quasi-random number generation and for advice on their usage.

\end{document}